# Uncovering Hidden Bias in Neutron Diffraction Residual Strain Measurements

Cole Franz,[a] Michael B. Prime,[b] Jeffrey Bunn,[c] Andrew Payzant,[c] Katharine Page[a]

[a] *Tickle College of Engineering, Knoxville, TN, USA.*

[b] *Los Alamos National Laboratory, Los Alamos, NM, USA.*

[c] *Oak Ridge National Laboratory, Oak Ridge, TN, USA.*

---

When calculating residual strain via neutron or X-ray diffraction, uncertainties propagated from the peak fit are often inadequate to describe the true scatter of measurements about a singular strain state, such as one that should describe a macroscopic continuum. Because diffraction is inherently a selective process, orientation dependent scatter arises from the sub-sampling of strong microstructure and strain gradients. This paper investigates the appropriateness of propagated uncertainties with reference to their original intention, i.e., noise about a mean value. Thirty-six unique orientations of strain measurements are taken at multiple locations within an additive friction-stir deposition component with fine-scale gradients (~200 μm) of plastic strain, texture, and residual elastic strain. Multiple strain and stress calculation pathways are compared: direct substitution of three measurements into Hooke's law, direct inversion of any six unique orientations into the strain state tensor, and thirty-six measurement least-squares estimation. For the latter two cases, the appropriateness of the uncertainty interval is statistically evaluated based on a physical constraint: common agreement under the strain transformation law. For this sample, the direct inversion of six measurements retains a conservative estimate of the uncertainty. However, propagated uncertainties in the least-squares solution greatly underestimate the true experimental scatter. A simple pathway to estimate appropriate uncertainty intervals is suggested. These results demonstrate that interpretation of uncertainty in residual strain is strongly dependent on intrinsic, sample-dependent effects, and that oversampling orientations and statistical analysis can give more accurate results with realistic uncertainties.

---

## Introduction

Residual stress calculations made with neutron diffraction are central to qualifying advanced components, including additively manufactured (AM) materials. For these calculations, uncertainties are typically reported as one-standard-deviation intervals derived from peak fitting. When propagated through the strain or stress tensor, these uncertainties are expected to describe the variability of the solution. For example, two independently computed solutions should agree within their one-standard-deviation intervals approximately 68% of the time. In practice, however, agreement is often significantly poorer, particularly in AM components that exhibit steep gradients in elastic strain, stored plastic strain, and crystallographic texture.

[1]This manuscript has been authored by UT-Battelle, LLC under Contract No. DE-AC05-00OR22725 with the U.S. Department of Energy. The United States Government retains and the publisher, by accepting the article for publication, acknowledges that the United States Government retains a non-exclusive, paid-up, irrevocable, world-wide license to publish or reproduce the published form of this manuscript, or allow others to do so, for United States Government purposes. The Department of Energy will provide public access to these results of federally sponsored research in accordance with the DOE Public Access Plan (http://energy.gov/downloads/doe-public-access-plan).

This discrepancy reflects a broader issue common to many inverse problems: propagated measurement uncertainties do not necessarily capture the true variability of the solution. In residual stress analysis, inverse methods are used at multiple stages, including diffraction peak fitting and tensor reconstruction. It is well established that inverse problems may contain inherent and often unknowable bias, particularly when regularization or simplifying assumptions are required. For example, incremental relaxation methods introduce bias through smoothing, leading to uncertainty estimates that underestimate the true error (Beghini & Grossi, 2024; Beghini *et al.*, 2023). In such cases, the inadequacy of uncertainties obtained by simply propagating measurement uncertainties through the analysis has long been recognized (Prime & Hill, 2005). Diffraction methods also rely on inverse procedures, most notably the fitting of diffraction peaks to assumed peak shapes. In neutron diffraction, it is conventional to propagate uncertainties based on the counting statistics of a peak-fit. However, an experimental study comparing measurements from multiple neutron instruments on the same welded specimen found that these propagated uncertainties were "significantly lower than the actual uncertainties" (Akrivos *et al.*, 2020).

In neutron diffraction, uncertainty underestimation arises from both measurement-related and analysis-related effects. During measurement, systematic deviations may arise from peak-fitting assumptions, instrument alignment, or more fundamentally, from the interaction between the neutron beam and populations of aligned crystallites. In samples with strong texture or attenuation gradients, the ensemble of diffracting crystallites may not represent the geometric center of the measurement volume, leading to anomalous peak shifts. These effects can increase point-to-point scatter well beyond the nominal peak-fit uncertainty (Holden et al., 2015; Hutchings et al., 2005). While experimental strategies such as scanning detectors can mitigate some of these effects, a more fundamental issue remains. Specifically, diffraction measurements do not always sample a true average of the underlying strain continuum. Because neutron diffraction probes relatively large volumes (~ $mm^3$), measurements in materials with fine-scale gradients often represent only a partial sampling of the strain field. In such cases, the fitted peak position may deviate from the true volumetric average, not because of measurement noise, but because different orientations sample different subsets of the microstructure. This unrepresentative volume averaging is exacerbated by texture and attenuation gradients and reflects the absence of a representative volume element (RVE) within the measurement volume (Şeren *et al.*, 2023). As a result, the variability between measurements may be dominated by physical heterogeneity rather than random noise. This concept is captured in Figure 1. Thus, the residual strains should be sampled in a manner that acknowledges the variability and self-equilibration of the strain continuum, which occurs at increasingly smaller length-scales (Withers & Bhadeshia, 2001*a*,*b*).

Additional complications arise during the strain and stress tensor calculation. Since the tensor is reconstructed from a limited number of directional measurements, the problem is inherently inverse and may be poorly conditioned. Small deviations in the measured strains can therefore produce large variations in the calculated tensor components. A common approach to mitigate this instability is to combine many measurements in a least-squares framework, which reduces variance by averaging. This was initially done by Winholtz and Cohen and followed up by Bunn to understand the affects of multi-axial loading (Winholtz & Cohen; Bunn *et al.*, 2014). While effective at reducing scatter, this approach implicitly assumes that all measurements represent a single underlying strain state and does not account for systematic or individual variability among them. As a result, the propagated uncertainties from least-squares estimation may underestimate the true variability of the measurements.

Bias cannot be meaningfully evaluated unless the measurements are checked against a physical constraint. Mathematically, biases introduced by inverse solutions are generally unknowable unless additional physical constraints on the true solution can be applied (Beghini *et al.*, 2023). In neutron diffraction measurements, however, such a constraint exists. Since stress and strain are tensors, that constraint is the coordinate transformation law, which all measured lattice strains must satisfy if they represent components of a single strain tensor. In other words, the measured strains are not independent but must satisfy the strain transformation equation, often illustrated through Mohr's circle. A previous study on an aluminum forging demonstrated that neutron peak-fit uncertainties had to be more than doubled for the measured strains to agree with the transformation equation within uncertainty (Prime *et al.*, 2014). That study involved only limited redundant measurement, leaving open questions about how broadly this issue applies.

In this work, we investigate these effects using a representative component produced by solid-state AM, which exhibits steep gradients in texture, plastic strain, and elastic strain (≤ 200 μm) (Metz *et al.*, 2023, 2024). Three common stress calculation pathways are examined: (1) direct substitution of three orthogonal strain measurements into Hooke's law, (2) inversion of six unique orientations using the strain transformation equations, and (3) least-squares estimation of the strain tensor using thirty-six measurements. In this study, the peak-fit uncertainties are treated as accurate representations of measurement precision, but not necessarily of the variability of the underlying strain field. The results show that these one-standard-deviation peak-fit uncertainties behave as intended when propagated through the six-measurement direct inversion, capturing the empirical spread of solutions. In contrast, when many measurements are combined in a least-squares estimator, the propagated uncertainties converge to values that are no longer representative of the observed variability. This occurs because the assumption of a single strain state is violated, even though the transformation law itself remains valid. These findings demonstrate that, in materials with fine-scale heterogeneity, residual strains obtained by diffraction and represented as continuum strains must be interpreted with care and given sufficiently conservative uncertainties. The estimation of sufficient uncertainty depends strongly on both the measurement strategy and the analysis method.

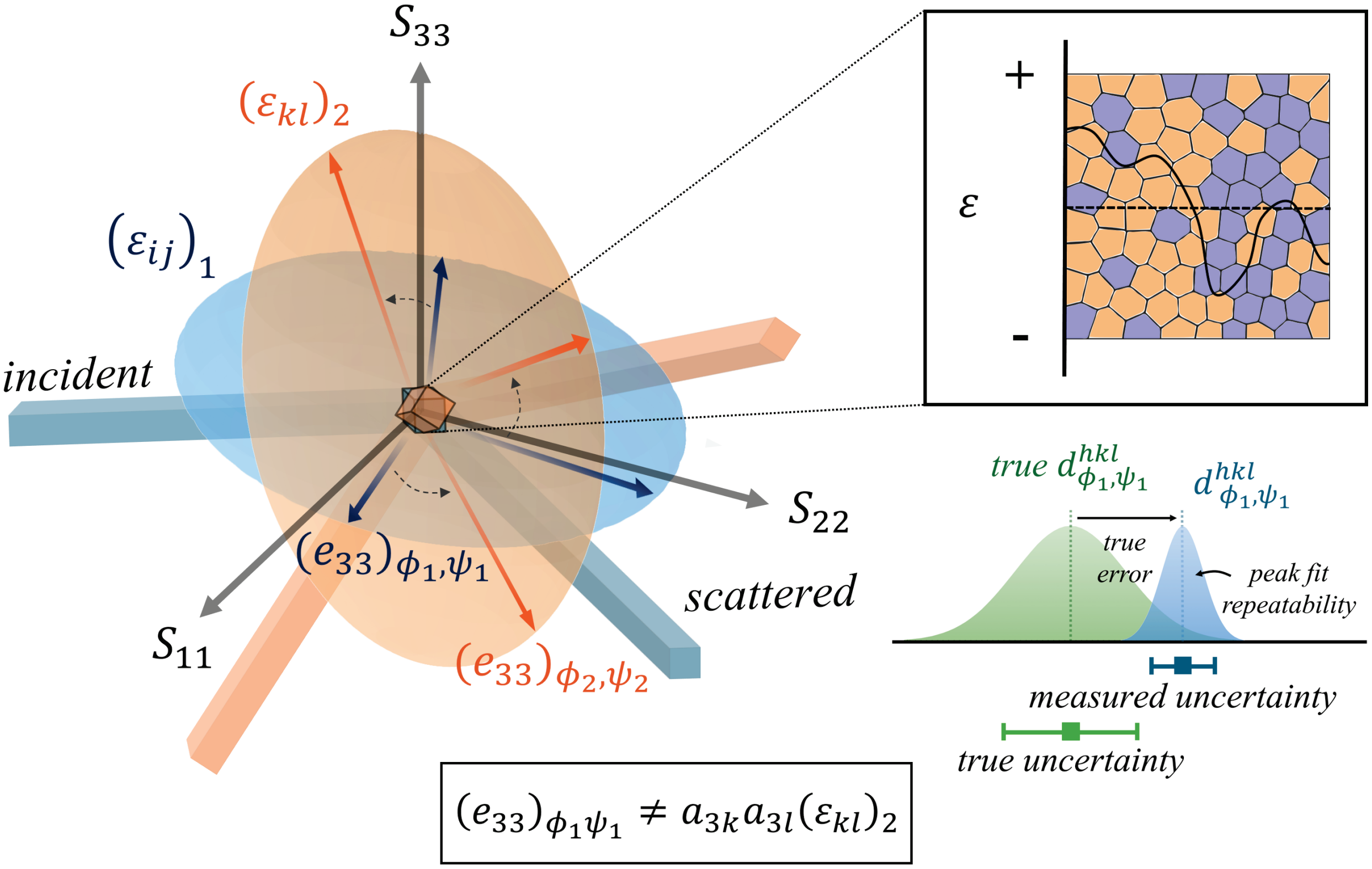

**Figure 1:** Schematic showing a case where two complimentary strain measurements produce unrepresentative volume averages, due to sampling of unique strain states within volumes of strong texture and strain gradients. The incident and scattered beam, each orientation of measured strain, and its sampled strain state (shown as strain ellipsoids) are separated by blue and orange colors. A true volumetric average, and its associated uncertainty, is more closely obtained by ensuring compatibility of measurements under the strain transformation law.

## Methods

### *AFSD Deposition Process*

Extruded AA6061-T6 bar stock (12.7 x 12.7 mm$^2$) was deposited onto a rolled AA6061-T6 baseplate (152.4 × 152.4 × 9.5 mm$^3$) using an L3 discrete feed additive friction-stir deposition (AFSD) machine (MELD Mfg., Christiansburg, VA, USA). In total, 11 unidirectional layers were deposited. Additional process specifications are given in the supplemental information. A diagram of the deposition procedure and AFSD coordinate system is given in Figure 2.

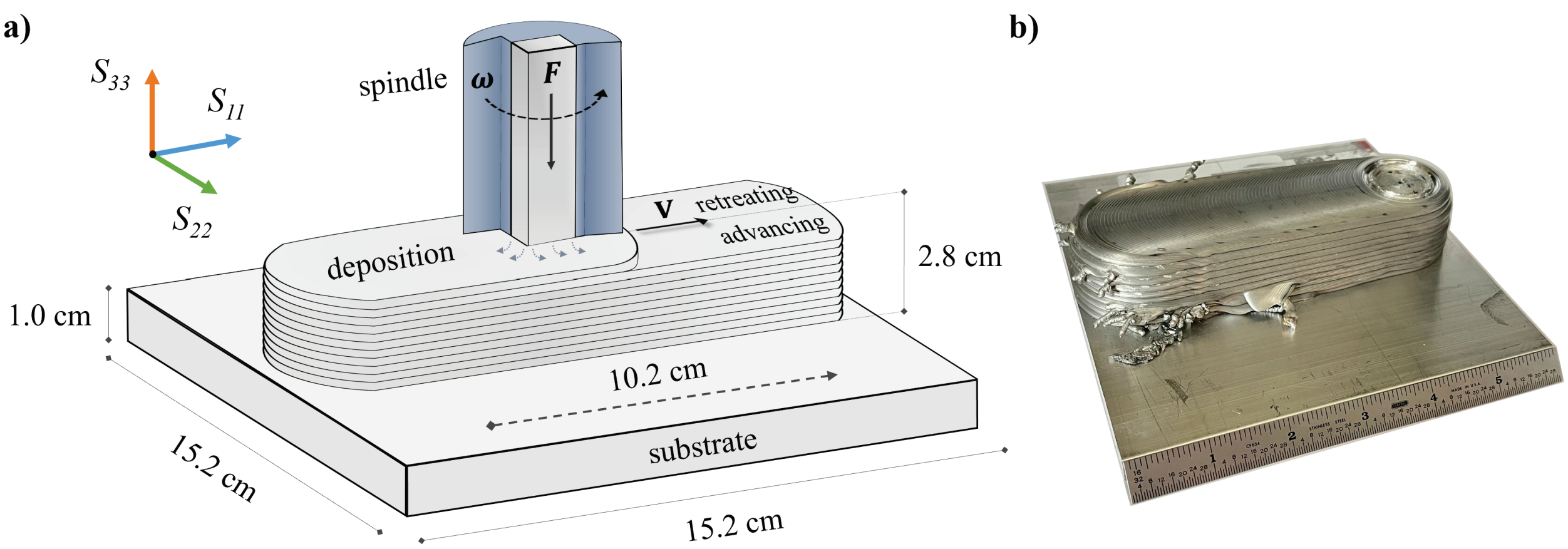

**Figure 2:** (a) Process diagram and coordinate system for AFSD. (b) Photograph of the AFSD component.

### *Neutron Diffraction*

Neutron diffraction was performed at the High Intensity Diffractometer for Residual Stress Analysis (HIDRA) at the High-Flux Isotope Reactor (HFIR), located at Oak Ridge National Laboratory (Bunn *et al.*, 2023). The incident and scattering slits were set up to isolate a 3 × 3 × 3 mm$^3$ volume in the AFSD sample. The scattering wavelength ($\lambda$ = 1.54 Å) was selected to probe the FCC-Al (311) interplanar spacing with a near isotropic gauge volume ($2\theta \sim 78°$). Seven locations, each approximately 2.54 mm apart (i.e. AFSD deposition layer height) were sampled, starting from the top surface ($S_{33} = 0$) and moving down two-thirds of the component height ($S_{33}$: -2.80 mm, -5.21 mm, -7.62 mm, -10.16 mm, -12.70 mm, -15.24 mm, -17.78 mm). At each location, 36 orientations of strain measurements were taken over one-quarter of the total orientation space. The peaks were fit with pseudo-Voight models using the *pyRS* analysis package (Fancher *et al.*, 2021). The diffraction and sample coordinate systems and angular conventions are shown in Figure 3, alongside a photograph of the experimental setup. The in-plane and out-of-plane angle pairs $(\phi, \psi)$ for each sampled orientation are given in Table 1. The signal was counted for 10 minutes to achieve an average fitted peak position uncertainty equivalent to ± 50 – 100 με.

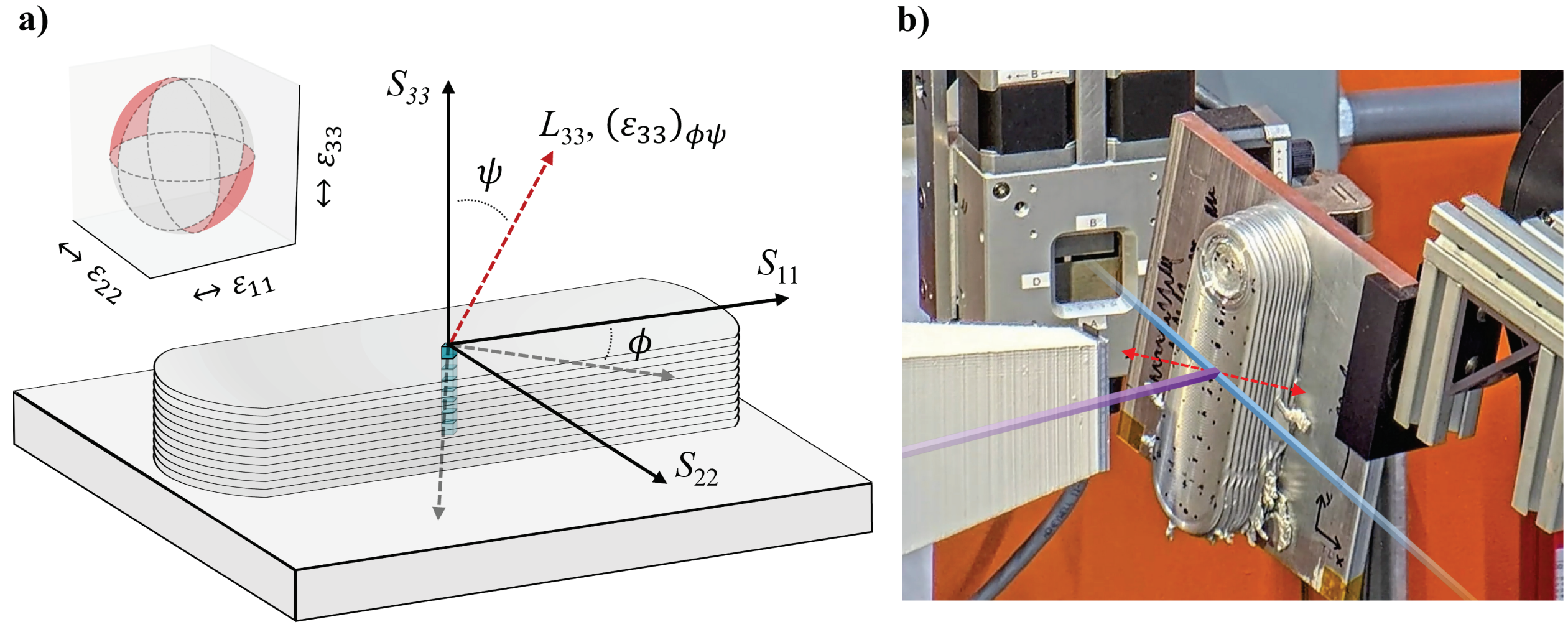

**Figure 3:** (a) Convention for sample and diffraction coordinate systems. Top-left inset shows the region of strain-space sampled ($0° < \phi < 90°$, $90° < \psi < 180°$). Each approximately isometric gauge volume is highlighted in blue. (b) Photograph of the AFSD sample within the neutron diffraction setup, which consists of multiple rotation stages mounted in tandem to rotate the sample about the center of each gauge volume. The transmitted beam, diffracted beam, and the diffraction (strain) vector are shown in blue, purple, and red, respectively.

**Table 1:** Sampled orientations of lattice parameters.

| | $\phi$ (°) | $\psi$ (°) |
|---|---|---|
| $d^{(\phi,\psi)}$ | 0 | 90, 105, 120, 135, 150, 180 |
| | 15 | 90, 105, 120, 135, 150 |
| | 45 | 90, 105, 120, 135, 150 |
| | 60 | 90, 105, 120, 135, 150 |
| | 75 | 90, 105, 120, 135, 150 |
| | 90 | 90, 105, 120, 135, 150 |
| $d_0^{(\phi,\psi)}$ | 0 | 90 |
| | 45 | 90 |
| | 69 | 90 |
| | 90 | 90 |

It is assumed that a strain-free reference was achieved by mechanical relief. For this purpose, a 5 mm diameter pillar was sectioned from an identical build, covering the same volumes as those sampled in the un-sectioned component (see supplemental information, Figure S1). To investigate sample texture, pole figures were collected from the (220), (311), (222), and (331) reflections using the strain-free reference pillar mounted on a miniature eulerian cradle. Texture was only collected at the bottom-most and top-most volumes ($S_{33}$: -2.80 mm, -17.78 mm). Orientation distribution functions (ODFs) were reconstructed from the measured pole figures using established conventions in the MTEX toolbox for MATLAB (Bachmann *et al.*, 2010; Hielscher & Schaeben, 2008).

*Strain and stress tensor calculation*

In the following passages, three different methods of calculations are compared. Each method is named according to the number of independent strain measurements used (e.g., 3-strains, 6-strains, or 36-strains). The conventions for stress, standard deviation, and uncertainty are written as $\sigma_{ij}$, $s$, and $\mu$, respectively.

The measured strains are first calculated based on displacement of the (311) interplanar spacing with respect to its reference value,

$$(\varepsilon_{33})_{\phi\psi} = \frac{d^{\phi\psi} - d_0}{d_0} \tag{1}$$

The total variation in $d_0$ for the (311) reflection was equivalent to ± 87 με, which is quite small. Thus, an average $d_0$ was used for all strain calculations in this paper.

<u>Method #1 – Three-measurements</u>

The most common method of residual stress mapping involves substitution of three mutually orthogonal strains into Hooke's Law for isotropic elasticity,

$$\sigma_{ii} = \frac{E}{1-v}\left(\varepsilon_{ii} + \frac{v}{1-2v}[\varepsilon_{11} + \varepsilon_{22} + \varepsilon_{33}]\right) \tag{2}$$

These strains are often measured along the component axes, but the equation is valid for any orthogonal set of directions. To calculate stress, strains measured along orientations $(\phi, \psi)$ parallel to the three sample axes: $S_{11}$, (0°, 90°); $S_{22}$, (90°, 90°); $S_{33}$, (0°, 180°), are substituted into Eq. 2. In this dataset, there is only a single instance of this solution for each depth. The elastic constants $E$ = 70 GPa and $v$ = 0.35 were used based on the elastic behavior predicted from single crystal properties and the sample ODFs (Hielscher & Schaeben, 2008; Bachmann *et al.*, 2010) (see supplemental information, Figure S2).

Method #2 – Six-measurements

The previous stress calculation is perfectly valid for isotropic elasticity even if there are shear strains. Though, if the principal components of strain / stress are misaligned from the three chosen orientations, Eq. (2) will not capture the potentially significant influence of the shear stresses on mechanical behavior. To obtain all six independent components of the strain tensor directly, any six strains measured in the diffraction coordinate system $(\varepsilon_{33})_{\phi\psi}$ can be used to calculate the sample strain state in the sample coordinate system. These will be referred to as the "sample strains." First, the sample strains are expressed in terms of the measured strains using the fundamental law of strain transformation:

$$(\varepsilon_{33})_{\phi\psi} = \varepsilon_{11}\cos^2\phi\sin^2\psi + \varepsilon_{22}\sin^2\phi\sin^2\psi + \varepsilon_{33}\cos^2\psi$$
$$+\varepsilon_{12}\sin 2\phi\sin^2\psi + \varepsilon_{13}\cos\phi\sin 2\psi + \varepsilon_{23}\sin\phi\sin 2\psi \quad (3)$$

Computationally, this system of equations is solved using linear algebra. In this case, Eq. 3 in matrix notation is written as,

$$\boldsymbol{e}_i = \boldsymbol{f}_j(\phi_i, \psi_i) \cdot \boldsymbol{\varepsilon}_j \quad (4)$$

Where $\boldsymbol{e}_i$ is an 6 × 1 vector of measured strains $(\varepsilon_{33})_{\phi\psi}$, $\boldsymbol{f}(\phi_i, \psi_i)$ is a 6 × 6 matrix of transformation coefficients (i.e. trigonometric functions of $\phi$ and $\psi$), and $\boldsymbol{\varepsilon}_j$ is 6 × 1 vector containing the sample strains (i.e. the components of the strain tensor). The sample strains are found by matrix inversion:

$$\boldsymbol{f}_j(\phi_i, \psi_i)^{-1} \cdot \boldsymbol{e}_i = \boldsymbol{\varepsilon}_j \quad (5)$$

This direct solution requires that the six chosen measurement orientations be linearly independent; otherwise, the system is ill-posed due to correlation among measurements. In practice, the numerical conditioning of the transformation matrix $\boldsymbol{f}_j(\phi_i, \psi_i)^{-1}$ determines how strongly measurement uncertainties are amplified in the recovered tensor components. Well-spaced orientations yield a well-conditioned system, whereas nearly redundant orientations can lead to unreasonable solutions and uncertainties.

To demonstrate the effects of numerical conditioning, all possible strain tensor components were calculated using unique combinations of six strain measurements drawn from the available set of thirty-six,

$$\frac{n!}{k!(n-k)!} \quad (6)$$

where $n$ = 36 is the total number of measured strains at a given location and $k$ = 6 is the number of strains required for a direct tensor inversion. This yields 1,947,792 unique combinations per location, of which approximately 0.8% are non-invertible owing to linear dependence within $\boldsymbol{f}_j(\phi_i, \psi_i)^{-1}$. For each valid combination, the full sample strain tensor was recovered by solving the linear system in Eq. 5. Since the number of feasible solutions is exceedingly large, visualization was performed by randomly subsampling 50,000 solutions per component and estimating their continuous distributions using an adaptive, weighted kernel density estimation (KDE):

$$\hat{f}(x) = \frac{1}{\sum_{i=1}^{n} w_i h_i'} \sum_{i=1}^{n} w_i K\left(\frac{x - x_i}{h_i}\right) \tag{7}$$

where $K$ is a Gaussian kernel, $w_i = 1 / \mu_i^2$ are inverse-variance weights based on measurement uncertainty, and $h_i'$ are local adaptive bandwidths containing the measurement uncertainty,

$$h_i' = \sqrt{h_i^2 + \sigma_i^2} \tag{8}$$

where $h_i$ is the initial density estimate using Abramson's rule.

Hooke's law for isotropic elasticity was applied to calculate stress,

$$\sigma_{ij} = \frac{E}{1+v} \varepsilon_{ij} + \frac{vE}{(1+v)(1-2v)} \delta_{ij} \varepsilon_{kk} \tag{9}$$

where $\delta_{ij}$ is the Kronecker delta. The resulting stress tensor components were likewise visualized using adaptive KDEs, with preprocessing steps (outlier rejection, normalization, and adaptive bandwidth selection) identical to those used for the strain distributions. Outlier removal for all analyses is based on the interquartile method, which removes points beyond 1.5× the interquartile range from the first and third quartiles.

Method #3 – Thirty-Six measurements

To incorporate all 36 strain measurements simultaneously, a weighted least-squares formulation was adopted following Winholtz and Cohen (Winholtz & Cohen). The closed-form solution is expressed as,

$$\boldsymbol{\varepsilon}_j = \left[\boldsymbol{f}_j(\phi_i, \psi_i)^T \, \boldsymbol{W} \, \boldsymbol{f}_j(\phi_i, \psi_i)\right]^{-1} \boldsymbol{f}_j(\phi_i, \psi_i)^T \, \boldsymbol{W} \, \boldsymbol{e}_i \tag{10}$$

where $\boldsymbol{W}$ is the diagonal weight matrix, $\boldsymbol{W} = \mathrm{diag}\left(\frac{\boldsymbol{1}}{\boldsymbol{\mu}_{e_i}^2}\right)$, and $\boldsymbol{e}_i$ is an 36 × 1 vector of measured strains $(\varepsilon_{33})_{\phi\psi}$, $\boldsymbol{f}(\phi_i, \psi_i)$ is a 36 × 6 matrix of transformation coefficients (i.e. trigonometric functions of $\phi$ and $\psi$), and $\boldsymbol{\varepsilon}_j$ is 6 × 1 vector containing the sample strains (i.e. the components of the strain tensor). This approach assigns greater influence to measurements with smaller uncertainty. For reference, the detectability limit of the HIDRA beamline is equivalent to ± 100 to 150 με.

*Uncertainty propagation*

Uncertainties were evaluated following ISO/IEC Guide 98-3:2008 (GUM) (JCGM GUM-6:2020 Guide to the expression of uncertainty in measurement-Part 6: Developing and using measurement models Guide pour l'expression de l'incertitude de mesure-Partie 6: Élaboration et utilisation des modèles de mesure, 2020). The uncertainties in each measured strain were obtained by the law of propagation of uncertainty,

$$u^2(\varepsilon_{hkl,i}) = \left(\frac{\partial \varepsilon}{\partial d_i}\right)^2 u^2(d_i) + \left(\frac{\partial \varepsilon}{\partial d_{0,i}}\right)^2 u^2(d_{0,i}) + 2\,\frac{\partial \varepsilon}{\partial d_i}\frac{\partial \varepsilon}{\partial d_{0,i}}\,\mathrm{cov}(d_i, d_{0,i}) \tag{11}$$

Under the working assumption that $d_i$ and $d_{0,i}$ are independent,

$$u(\varepsilon_{hkl,i}) = \sqrt{\frac{u^2(d_i)}{d_{0,i}^2} + \frac{d_i^2}{d_{0,i}^4}\,u^2(d_{0,i})} \tag{12}$$

the uncertainty of the estimated tensor components followed from the covariance of $\boldsymbol{\varepsilon}_j$. With uncorrelated measurements,

$$\mathrm{cov}(\boldsymbol{\varepsilon}) = (\boldsymbol{f}_j(\phi_i, \psi_i)^{\top}\boldsymbol{W}\boldsymbol{f}_j(\phi_i, \psi_i))^{-1} \tag{13}$$

thus, the component standard uncertainties are the square roots of the diagonal elements:

$$u(\varepsilon_{ij}) = \sqrt{[\mathrm{cov}(\boldsymbol{\varepsilon})]_{\boldsymbol{kk}}} \tag{14}$$

## Results

### *Correlation of crystallite orientation and signal*

In Figure 4a, the measured signals for the (311) reflection are reported alongside predictions of the strength of the (311) crystallographic orientation. Each figure is constructed based on interpolation of the data within the breadth of the sampled orientation space $(\phi, \psi)$ onto an equal-area projection. Variations in strength of preferred orientation, reported as multiples of random distribution (MRD), are weak for the (311) plane, which is why it is typically recommended for strain measurement (Hutchings et al., 2005). Peak area (i.e. signal) is well correlated with crystallite orientation, and each of these metrics are inversely correlated with fitted peak uncertainty. Representative peak fits along the three sample axes are shown in Figure 4b. The (311) peak height-to-background ratio varies modestly across the measured orientation space, with values spanning roughly 0.44 – 2.94 (median ~ 1.0). A conservative estimate was made for the number of grains within each volume. If the grain size is assumed to be 20 µm (based on a similar alloy and deposition process (Metz *et al.*, 2024)) within the 3 × 3 × 3 mm$^3$ measurement volumes, the total number of grains within each volume is on the order of 3.4 × 10$^6$. Due to fine-scale gradients in texture, only some fraction of these grains participate in diffraction (Metz *et al.*, 2024).

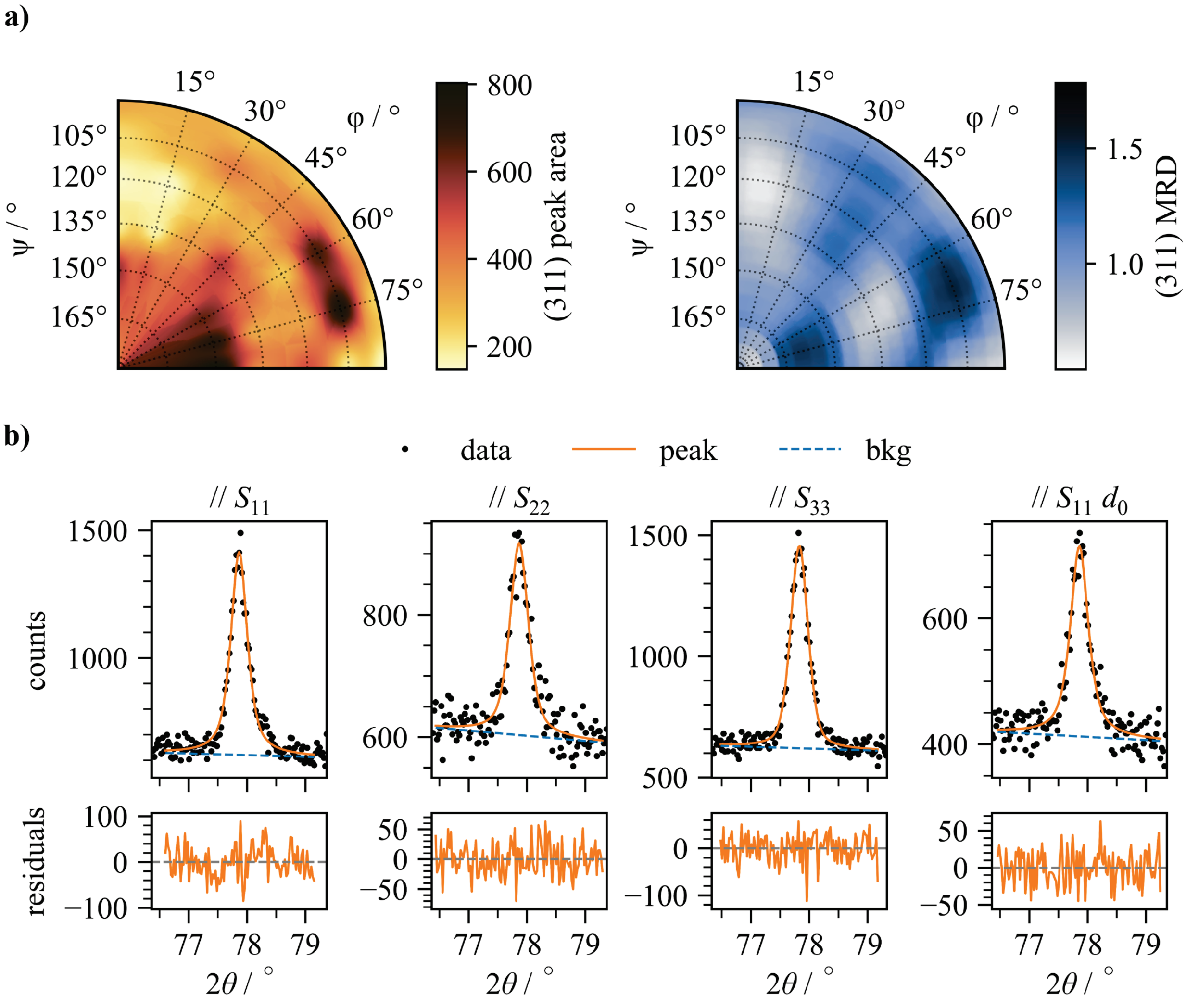

**Figure 4:** (a) The measured signal (left), computed as peak area, correlates well with the predicted crystallographic orientation of the (311) reflection (right), based on the ODF reconstruction (data are shown for the top volume, $S_{33}$ = -2.8 mm). (b) Raw data and pseudo-Voight peak fits from the top volume, shown along three orthogonal directions, including data for $d_0$.

*Variation of the reference interplanar spacing*

Figure 5 shows all measured (311) interplanar spacings. The total variation in interplanar spacing is approximately an order of magnitude larger than that of the reference values ($10^{-3}$ με versus $10^{-4}$ με equivalent). Accordingly, an average reference value of 1.21656 Å was adopted for all strain calculations. At the extreme, this assumption could incur an error of up to 164 με, corresponding to 10 MPa in annealed AA6061.

Using this fixed reference, most of the calculated lattice strains are tensile, consistent with prior neutron diffraction measurements in similar regions of friction-stir deposited AA6061 (Zhu *et al.*, 2023). In that study, the total variation in $d_0$ was equivalent to 700 με; however, that component was approximately twice the size of the present build and experienced roughly six times more depositions. Another investigation of friction-stir deposited AA7075 reported a total variation in $d_0$ equivalent to 300 με for a build of comparable geometry and number of depositions to this study (Metz *et al.*). The total possible variation in the (311) interplanar spacing, based on matrix solute concentrations—whether from solute supersaturation or precipitation—may vary up to 1400 με equivalent in AA6061 (see supplemental information, Figure S3). However, that is not the case here.

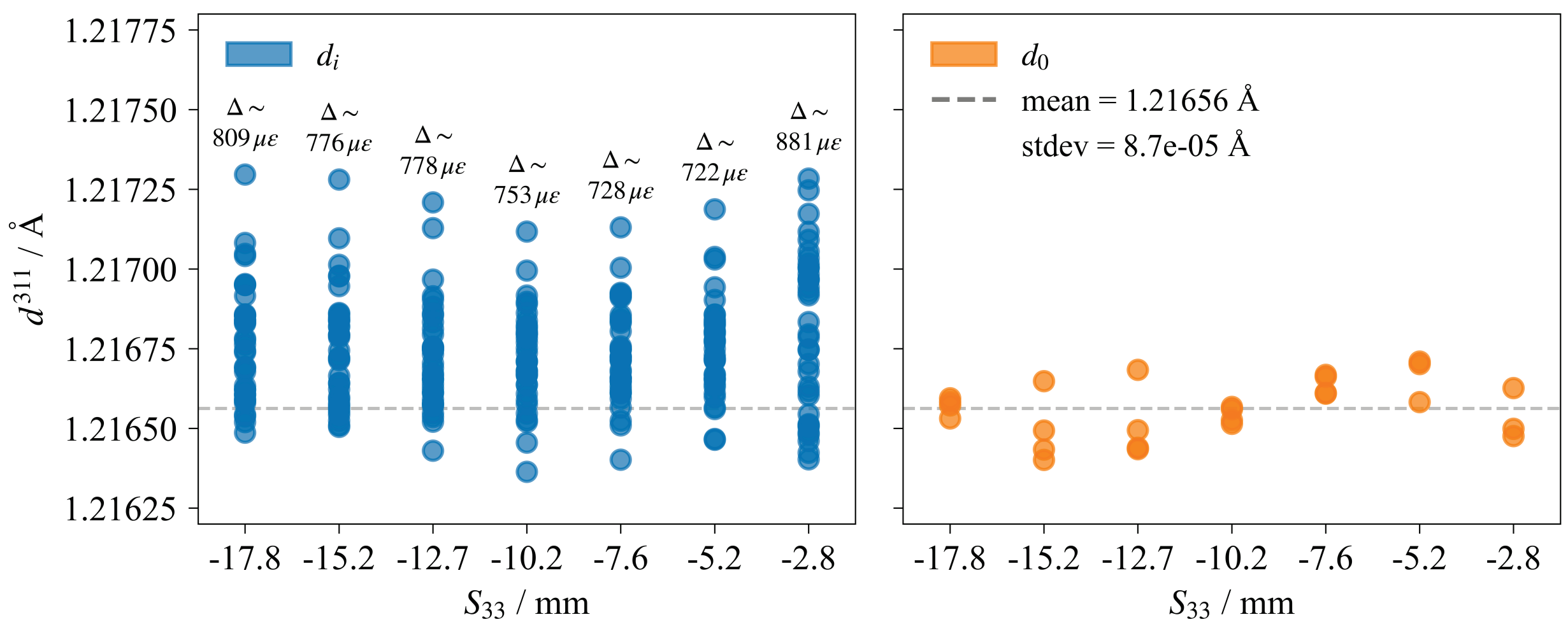

**Figure 5:** Scatterplots of all measured, sample lattice parameters (left) and reference lattice parameters (right). The total variation in sample lattice parameters, denoted above each location in microstrain equivalent, is an order of magnitude greater than the variation in the reference parameters. The average reference lattice parameter used in all calculations is shown with a dashed line.

*Verification of physical bias*

When calculating the strain tensor via the strain transformation law, it is assumed that all measurements represent the same strain-state. When that is not true, bias is incurred within each measurement based on its deviation from the true volumetric average of residual strain within the gauge volumes. The existence of physical biases within the strain measurements, primarily due to centroid misalignment and unrepresentative volume averaging, was investigated. Based on the available data, the relative influence of each type of bias is indistinguishable. However, the combined influence is evident by examining the behavior of a $\sin^2\psi$ plot. A $\sin^2\psi$ plot examines the relationship between diffraction geometry ($\sin^2\psi$) and the measured strains $(\varepsilon_{33})_{\phi\psi}$. For a single, representative stress state, the $\sin^2\psi$ plots exhibit either linear (biaxial stress state) or symmetric curvilinear behavior (triaxial stress state). The latter is called *psi-splitting (Noyan & Cohen, 1987)*. Oscillatory behavior implies that strong strain gradients are sampled, violating the RVE assumption and potentially compromising the accuracy of the data when checked against the strain transformation law. In addition to the $\sin^2\psi$ plot, the influence of strongly heterogenous intergranular strains was investigated. It was assumed that variations in fitted integral breadth (peak area divided by height) were positively correlated with variations in intergranular strains. The results of the $\sin^2\psi$ and integral breadth tests are shown in Figure 6.

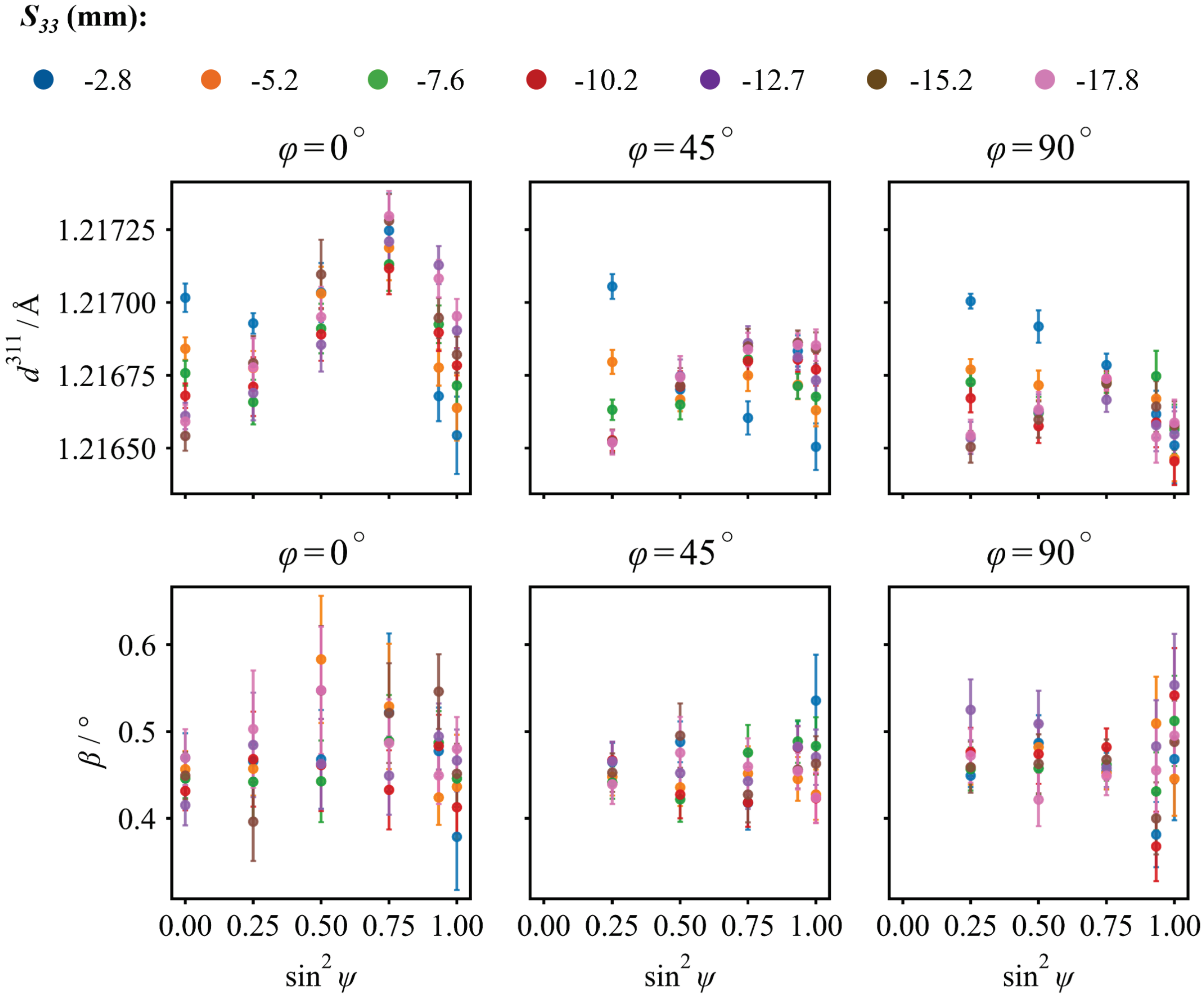

**Figure 6:** The top row shows three $\sin^2\psi$ tests with oscillatory behavior, suggesting that the measurements do not consistently sample a single representative strain state. The bottom row shows a strong influence of the measured orientation on the integral breadth, which is assumed to be correlated with intergranular strain variations. Both tests suggest that significant biases exists within the strain measurements, which if left unaccounted for, will strongly influence the calculated stresses and will lead to underestimated uncertainties.

Each test revealed strongly oscillatory behavior, indicating strong orientation dependence of the measured strain state. This verifies that some combined influence of centroid misalignment and unrepresentative volume averaging produce strong biases in the measured strains. These variations in integral breadth, and variations in crystallite orientation previously described, both influence peak shape, which together explain the variability in fitted peak uncertainties (± 50 to 100 με).

*Strain and stress tensor calculation*

Method #1 – Three measurements

Figure 7 shows the results from substituting three orthogonally measured strains directly into Hooke's Law. The average uncertainty in strain is ± 100 με, approximately equal to the reported detectability of the instrument (Bunn *et al.*, 2023). The average uncertainty in stress is 20 MPa.

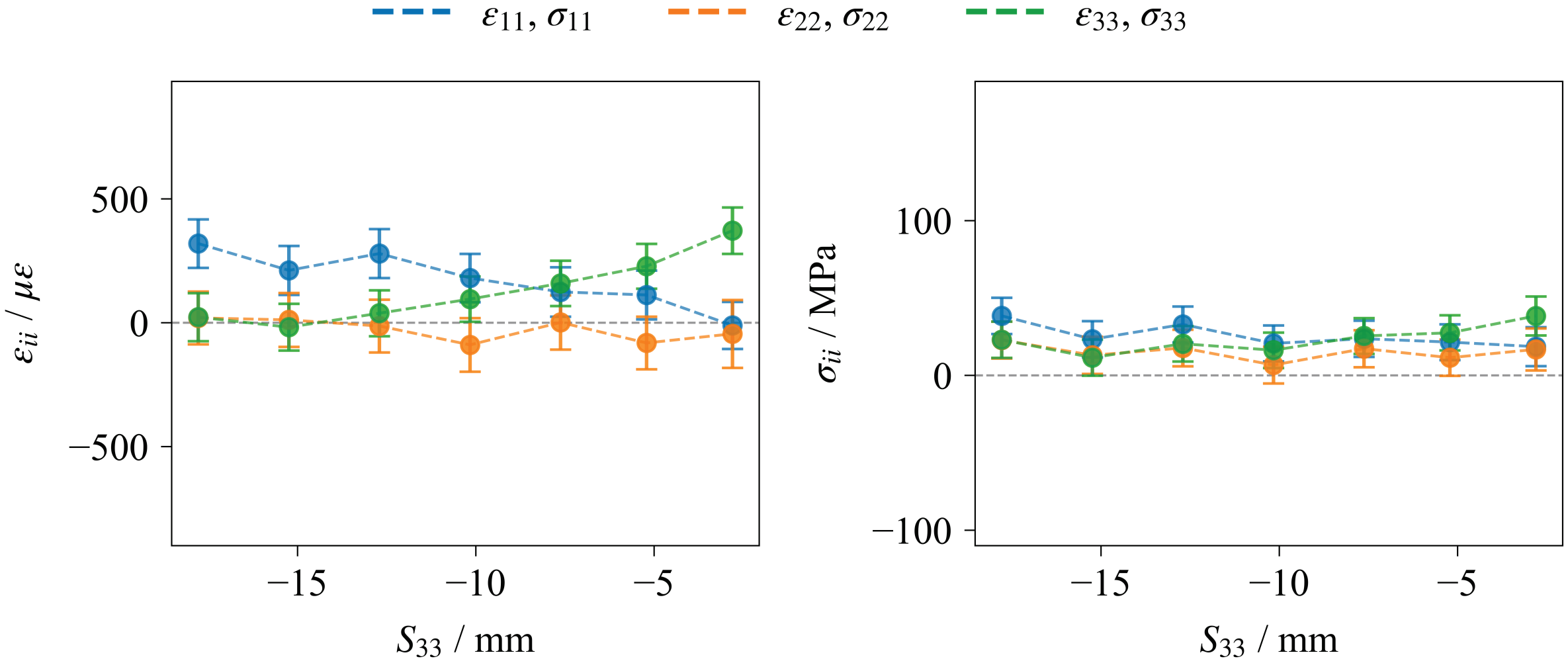

**Figure 7:** Strain and stress tensor results from the three-component analysis plotted against depth. The calculated strains and stresses are predominantly tensile.

Method #2 – Six measurements

Figure 8 shows continuous distributions of all calculable sample strain / stress tensor components, based on combinations of six orientations from thirty-six available measurements. The strains / stresses can be either tensile or compressive depending on the choice of orientations, spanning a range greater than 5× the 62 MPa yield strength of annealed or friction-stir deposited AA6061 (Anderson *et al.*, 2019; Rutherford *et al.*, 2020). Averaging the standard deviation of each distribution across the depths shows that the shear-strain components ($\varepsilon_{12}, \varepsilon_{13}, \varepsilon_{23}$) exhibit markedly larger dispersion than the normal-strain components $(\varepsilon_{11}, \varepsilon_{22}, \varepsilon_{33})$: 293 με versus 148 με (1.98×). In contrast, the shear-stress components $(\sigma_{12}, \sigma_{13}, \sigma_{23})$ are much less dispersed than the normal-stress components $(\sigma_{11}, \sigma_{22}, \sigma_{33})$: 26 MPa versus 103 MPa (0.26×).

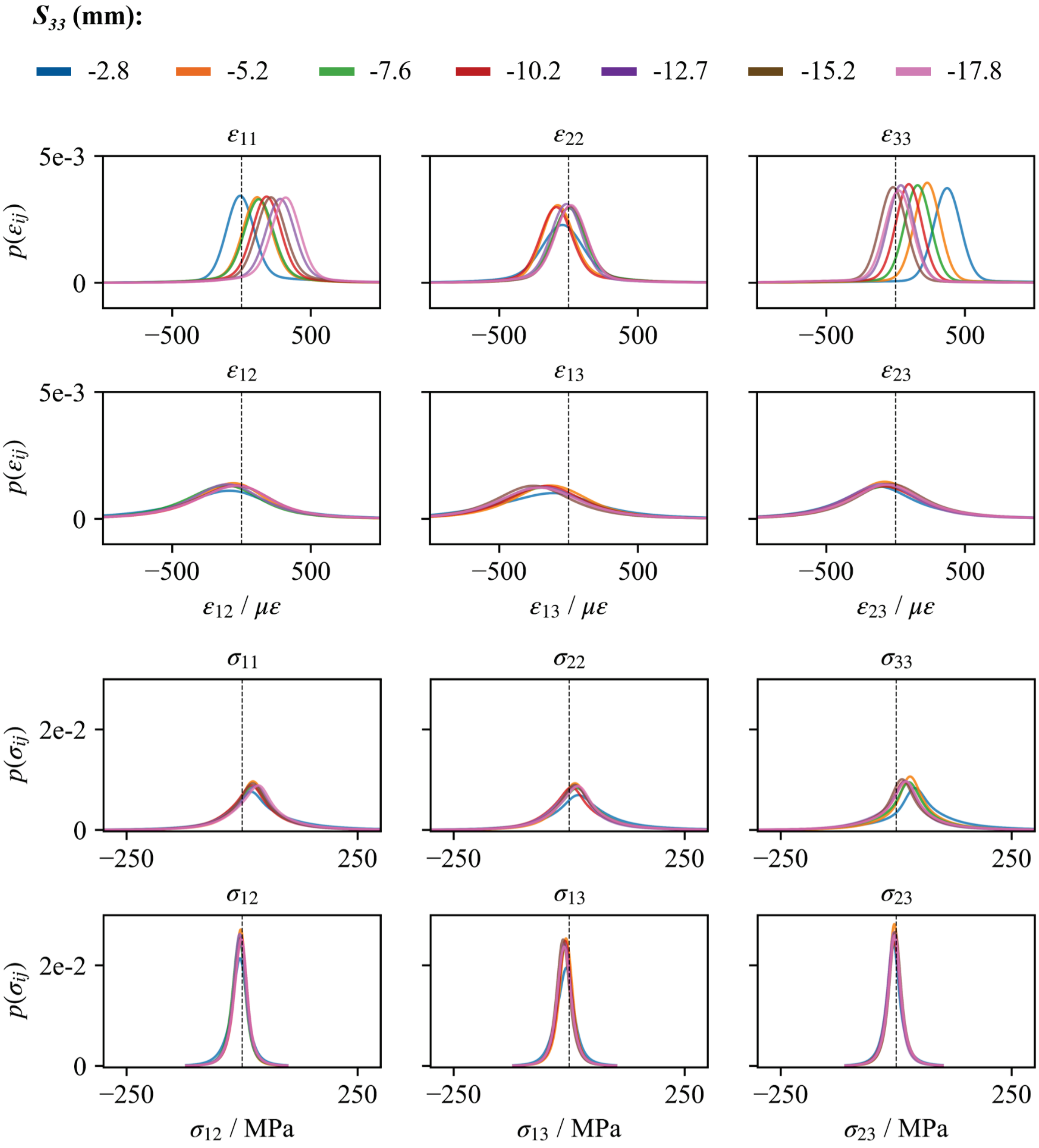

**Figure 8:** Continuous distributions (KDEs) of all calculable six-component solutions for strain (top) and stress (bottom) are extremely broad, indicating that taking a random combination of 6 orientations might not even get one the correct sign for strain or stress.

Two effects may explain these trends. Firstly, when solving the six-component inversion, the measurement operator that maps rotated lattice-spacing observations to sample-frame strains (i.e., $\boldsymbol{f}_j(\phi_i, \psi_i)^{-1}$) has uneven leverage across components. Many six-angle subsets provide strong constraints on the normal terms but weak constraints on the shear terms. In the strain transformation law, nearly coplanar or symmetrically clustered orientations make the corresponding columns of the design matrix $\boldsymbol{f}_j(\phi_i, \psi_i)$ partially collinear, amplifying noise and yielding broader KDEs for $\varepsilon_{\text{shear}}$. Second, the isotropic stress equation $\boldsymbol{\sigma} = \lambda\,\text{tr}(\boldsymbol{\varepsilon})\,\boldsymbol{I} + 2\mu\,\boldsymbol{\varepsilon}$ (with $\lambda$ and $\mu$, Lame's first and second parameters) mixes the volumetric strain into all three normal stresses, so uncertainty in $\text{tr}(\boldsymbol{\varepsilon})$ is broadcast to $(\sigma_{11}, \sigma_{22}, \sigma_{33})$ with weight $\lambda$, whereas shear stresses depend only on the shear strains via the factor $2\mu$. Because $\lambda$ is of the same order as $2\mu$ for aluminum, variance in $\text{tr}(\boldsymbol{\varepsilon})$ (and in how $\varepsilon_{11}, \varepsilon_{22}, \varepsilon_{33}$ partition that trace) inflates $\sigma_{\text{normal}}$ relative to $\sigma_{\text{shear}}$, even though $\varepsilon_{\text{shear}}$ itself is more weakly constrained. Together, these conditioning and propagation effects yield the observed pattern: large spread in $\varepsilon_{\text{shear}}$, modest spread in $\sigma_{\text{shear}}$ (set mainly by $2\mu$), and the largest spread in $\sigma_{\text{norm}}$ due to the added volumetric coupling.

These complex uncertainty interactions make this data set an ideal test case. If the underlying peak-fit uncertainties are accurate and representative, uncertainty propagation through the equations will correctly capture the uncertainty in calculated quantities such as stress. The average solution for each of the distributions is plotted in Figure 9. The uncertainties are calculated as the average uncertainties of the ten most common solutions. These calculated uncertainties are approximately an order of magnitude larger than those obtained from the three-measurement method, primarily due to noise amplification arising from the geometric conditioning of the design matrix for non-ideal sets of orientations.

In most six-measurement combinations, the selected orientations do not align with the sample axes; thus, the transformation required to recover the strains along these axes amplifies the uncertainties. To be clear, the uncertainties shown in Figure 9 represent the uncertainty of a single, typical six-measurement solution constructed from an arbitrary set of orientations. They do not represent a fundamental lower bound on the uncertainty of the strain tensor, nor do they imply that six measurements inherently produce larger uncertainty than three measurements. Rather, they reflect the fact that most randomly selected orientation sets are not optimally conditioned for resolving the tensor components along the sample axes. This distinction is important. If the three orthogonal (sample axis-aligned) strain measurements are included and supplemented with three additional orientations, the uncertainty in the normal strain components remains the same as the three-measurement method. In this case, the additional measurements do not degrade the uncertainty because the primary components are already directly measured, and the system is well-conditioned. However, such a set of ideal orientations represents only a small subset of all possible six-measurement combinations. For most orientation sets, the strain tensor must be reconstructed through a transformation from non-axis-aligned measurements, and the associated conditioning of the design matrix leads to amplification of uncertainty in the recovered components. Thus, the larger uncertainties observed in Figure 9 are primarily a consequence of measurement geometry rather than an inherent limitation of the six-measurement approach. The effects of measurement selection, geometric conditioning, and the appropriateness of the resulting uncertainty magnitudes are further investigated in the discussion.

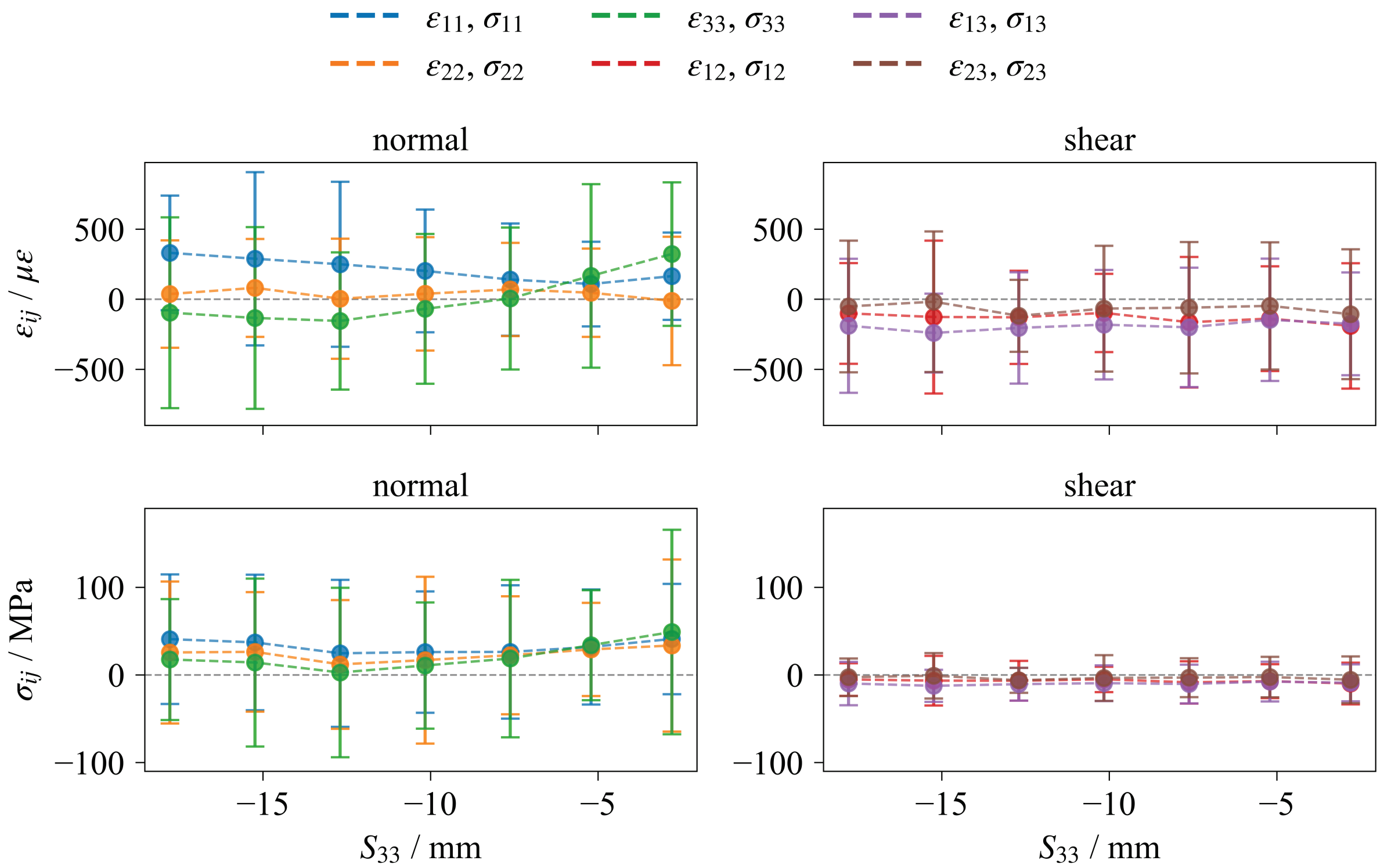

**Figure 9:** Results of the average solution for the strain and stress tensors (averages of the KDEs) given a choice of six orientations of strain measurements. The uncertainties are calculated as the average uncertainties of the ten most common solutions. Uncertainties for the six-measurement method are an order of magnitude larger than those of the three-measurement method, due to conditioning of the inversion process when obtaining sample strains.

<u>Method #3 – Thirty-Six measurements</u>

The least-squares procedure was used to incorporate all thirty-six strain measurements into a single solution (Eq. 10), the results of which are shown in Figure 10. As might be expected, the stress and strain values are very similar to those plotted in Figure 9, which are an average of the most common solutions for the six-measurement analysis. The propagated uncertainties are lower than those of the three- and six-measurement analysis by factors of 2× and 5×, respectively. Also, the point-to-point scatter in the curves is drastically reduced compared to the three-measurement analysis. This is the intended effect of incorporating many imprecise measurements, which decreases solution variance, yet potentially increases bias because the underlying variability in the strain continuum is not represented within the uncertainty intervals. The $\sin^2 \psi$ test clearly showed that the sampled strain states vary more strongly than these solutions suggest. This motivates the need to inflate the measurement uncertainties.

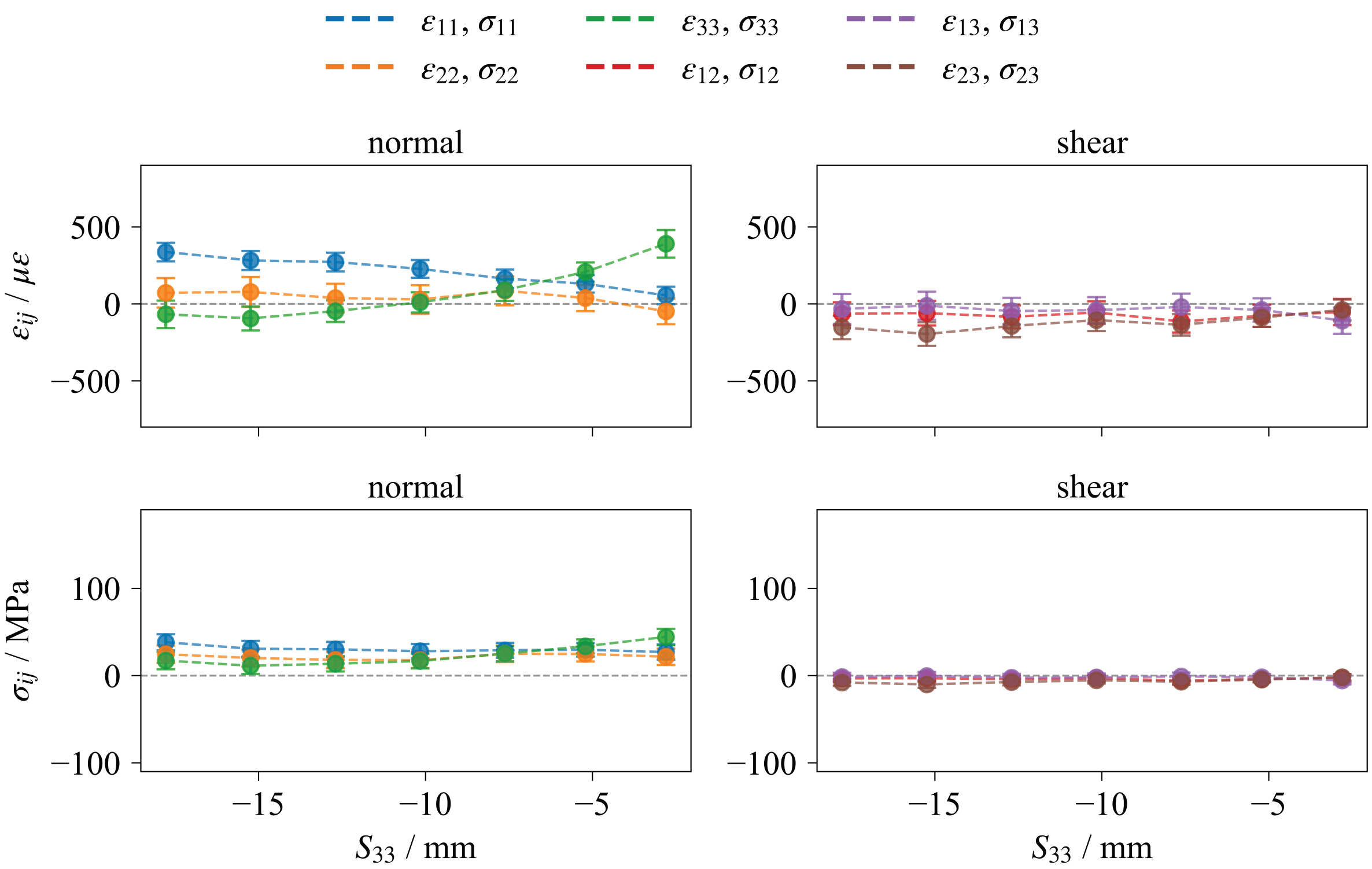

**Figure 10:** Results of the least-square analysis that includes all thirty-six measured strains. These profiles are smoother and with smaller apparent uncertainties than the preceding analyses. This likely indicates that strain / stress are now more closely resembling continuum quantities at the macro-scale.

**Discussion**

Based on the presence of bias in the measured strains, this discussion examines how uncertainty propagates to stress for both the six-measurement direct solutions and the thirty-six-measurement least-squares solution. The key distinction between these approaches lies in how their propagated uncertainties should be interpreted. It will be shown that the six-measurement method propagates uncertainties that retain the meaning of a one-standard-deviation measure of random noise. In contrast, the least-squares method produces a lower-variance estimate of the strain and stress state, but the propagated uncertainties no longer reflect the variability present in the original measurements. As a result, the interpretation of uncertainty differs between the two methods, particularly with respect to their relationship to the unknown ground truth for this dataset.

*Uncertainty analysis of direct solutions*

First, the six-measurement procedure was investigated to assess whether propagated one-standard-deviation uncertainties reflect the empirical spread of solutions. For this purpose, normalized deviations are formed for each tensor component *ij* and solution *i*,

$$D_{ij} = \frac{\varepsilon_{ij}^{(i)} - \overline{\varepsilon_{ij}}}{\mu_{\varepsilon_{ij}^{(i)}}} \tag{15}$$

where $\varepsilon_{ij}^{(i)}$ is the individual strain value, $\overline{\varepsilon_{ij}}$ is the mean of all strains, and $\mu_{\varepsilon_{ij}^{(i)}}$ is the uncertainty for that value. The same format is used for the stress deviation. This metric quantifies how far each solution lies from the mean relative to its own uncertainty. Values with magnitude less than one indicate agreement within one standard deviation, regardless of the absolute difference. This deviation metric is specifically designed to test whether the propagated uncertainties capture the observed spread of solutions. It does not penalize solutions with large uncertainties, provided those uncertainties are sufficient to explain the deviation from the mean. Importantly, this formulation does not assume that the mean represents the true volumetric strain state. Rather, the analysis is performed in a relative sense, which is appropriate because the strain transformation law constrains only relative values.

The means and standard deviation of the deviations are shown in Figure 11. If a standard deviation is greater than 1, the values are dispersed more broadly than predicted by the individual uncertainties, indicating those uncertainties are too small on average. The mean is harder to interpret. Since ground truth is not known for the strain state, a non-zero mean does not necessarily indicate a bias in the solutions. It only indicates that the deviations are systematically shifted to one side of the mean. Nonetheless, some interesting patterns appear in the means and are thus discussed.

Apart from the shallowest location, the uncertainty estimate is mildly conservative. Both the strain and stress components have standard deviations below 1, indicating propagated uncertainties that slightly exceed the spread of values. This result gives confidence in the peak fit uncertainties, a result not necessarily expected. At the shallowest depth, the standard deviation is predominately greater than zero with typical values around 1.2, which is investigated further shortly.

The normal strain components show consistently non-zero means – negative for $\varepsilon_{11}$ and $\varepsilon_{22}$ and positive for $\varepsilon_{33}$. These non-zero means are suppressed when mapped to stress. This sign structure may be compatible with a small systematic effect in the normal strains (e.g., the common $d_0$ or angular coverage) to which shear is insensitive and which is diluted by the isotropic Hooke relation through the $\lambda\, tr(\boldsymbol{\varepsilon})$ term. Overall, the stress deviations have means closer to zero than the strain deviations, suggesting that interpretation and reporting at the stress level are more robust to systematic effects from the measured strains and that carrying a shared $d_0$ may suppress some systematic effects.

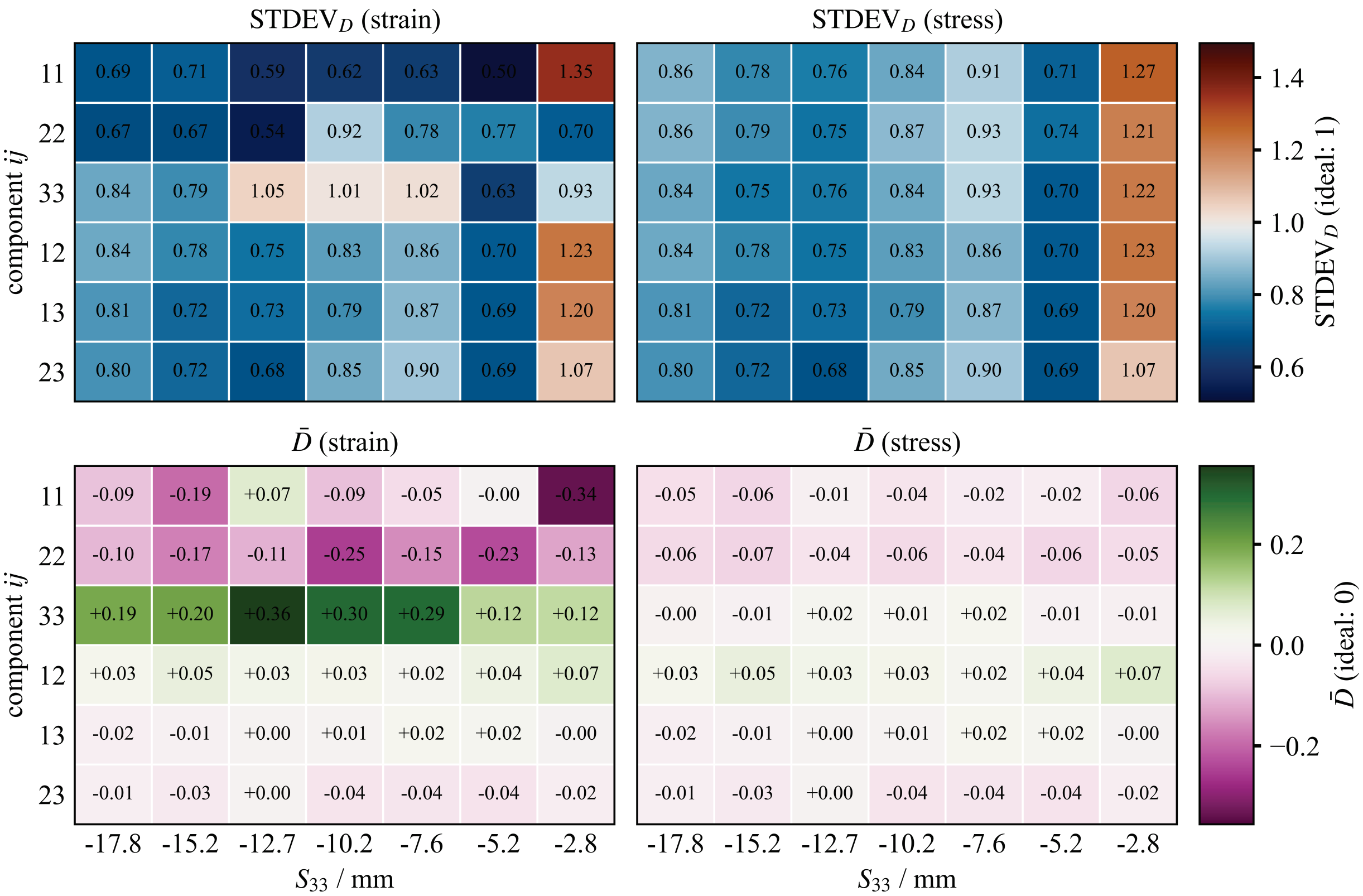

**Figure 11:** Statistical measures of systematic effects and the appropriateness of the propagated uncertainties, based on the mean and standard deviation of the normalized deviations to the six-measurement solutions (Eq. 15). Other than for the shallowest depth, the generally low values for the standard deviations indicate that the propagated peak-fit uncertainties give somewhat conservative estimates of the actual variance. Because the ground truths for the measured strains and stresses are unknown, it is difficult to assign significance to the deviation means significantly greater than zero.

To diagnose the physical and experimental drivers behind departures from the ideal normal distribution $D \sim \mathcal{N}(0,1)$ – particularly at the shallowest depth – each solved $\sigma_{ij}$ component is split into inliers ($|D| \leq 1$, where the deviation is close to 1, and outliers ($|D| > 1$), where the deviation is greater than 1. The following metrics are compared: (i) peak area (counting statistics), (ii) integral breadth (i.e. IB, fit stability/line broadening), (iii) a texture proxy (MRD mapped to $(\phi, \psi)$, only for $S_{33}$: -2.8 mm and -17.8 mm), and (iii) a geometry metric that summarizes angular coverage of the six directions used in each inversion (a proxy for conditioning of the design matrix $\boldsymbol{f}_j(\phi_i, \psi_i)$). Cellular maps, similar to Figure 11, are given in Figures S4 and S5 for each statistic across inliers and outliers.

It is assumed that maximizing the angular coverage of measurements leads to better conditioning of the design matrix. Thus, a geometry metric of primary interest is the minimum-spanning-tree (MST) angle sum for the six unit-directions $\{\hat{n}_k\}_{k=1}^{6}$:

$$S_{\text{MST}} = \sum_{(i,j)\in MST} \arccos\left(\left|\hat{n}_i \cdot \hat{n}_j\right|\right) \tag{16}$$

i.e., the total great-circle separation of the five edges that connect all six directions with minimum total length. This favors globally spread sets (better conditioning of the design matrix) while down-weighting clustered angles.

To test whether outliers differ systematically from inliers, we use an exact two-sided sign test on paired component-depth differences. For a given metric $M$ and for each component–depth cell $(c, d)$ (six components per depth), the paired difference is defined as

$$\Delta_{c,d} = \bar{M}_{outlier}(c, d) - \bar{M}_{inlier}(c, d) \tag{17}$$

Let $k$ be the number of positive $\Delta_{c,d}$ among $n$ non-tied pairs and assume the null of no systematic difference,

$$\text{H}_0: \mathbb{P}(\Delta > 0) = \mathbb{P}(\Delta < 0) = \frac{1}{2} \tag{18}$$

under which $k \sim \text{Binomial}(n, 0.5)$. The two-sided $p$-value is

$$p = 2\min\{\mathbb{P}(\text{K} \leq \text{k}), \mathbb{P}(\text{K} \geq \text{k})\} = 2\min\left\{\sum_{r=0}^{k}\binom{n}{r}2^{-n}, \sum_{r=k}^{k}\binom{n}{r}2^{-n}\right\} \tag{19}$$

This non-parametric test makes no normality assumption. It requires only approximate independence of the pairs across component-depth cells and symmetry of signs under $\text{H}_0$. With small $n$ (here $n$ = 6 within a depth; $n$ = 12 when pooling two depths, such as the top-most and bottom-most), $p$-values are discrete (e.g., the smallest two-sided $p$ at $n$ = 6 is 0.03125).

Table 2 pools both depths (12 pairs). Outliers have a slightly larger $S_{\text{MST}}$ on average that is statistically significant (+1.2%, $p$ = 0.0386). Peak area, IB, $d$-spacing uncertainty, and MRD all show very small average differences yet are not statistically significant. Table 3 resolves the comparison by depth: at the shallow location, outliers are weakly associated with lower area (−2.26%, $p$ = 0.031), higher IB (+0.57%, $p$ = 0.031), and higher $d$-spacing uncertainty (+2.58%, $p$ = 0.031). The MRD differences are uniformly small. Overall, there are some statistically significant differences between inliers and outliers, and between the top-most and bottom-most sampling locations, yet the actual magnitude of those differences is quite small (< 3%). The small differences between the top-most and bottom-most locations are likely due to the differences in thermal cycling. That is, the top-most layer has not undergone the additional recovery and recrystallization of the prior layers and thus retains a greater degree of accumulated plastic strain (i.e. integral breadth) and deformation texture strength (i.e. MRD).

**Table 2:** Cross-depth summary (both depths pooled; $n$ = 12 pairs).

| Metric | Mean Δ (out.−in.) | Rel. Δ (%) | Two-sided $p$ | Comments |
|---|---|---|---|---|
| Peak area (a.u.) | −3.48 | −1.0% | 0.774 | Small difference |
| Integral breadth (IB) | +0.0007 | +0.16% | 1.000 | No consistent difference |
| $d$-spacing uncertainty (με) | +0.50 | +1.0% | 1.000 | No consistent difference |
| Texture (MRD) | +0.002 | +0.17% | 0.774 | Small difference |
| $S_{MST}$ (°) | +1.58 | +1.2% | 0.0386 | Small cross-depth effect with consistent direction |

**Table 3:** Depth-resolved summary (component-paired within depth; $n$ = 6 pairs per depth).

| Metric | $S_{33}$ = -2.8 mm Δ (rel. %) | $p$ | $S_{33}$ = -17.8 mm Δ (rel. %) | $p$ |
|---|---|---|---|---|
| Peak area (a.u.) | −2.26% | 0.031 | +0.53% | 0.219 |
| Integral breadth (IB) | +0.57% | 0.031 | −0.26% | 0.031 |
| $d$-spacing uncertainty (με) | +2.58% | 0.031 | −0.65% | 0.031 |
| Texture (MRD) | −0.33% | 0.219 | +0.73% | 0.031 |
| $S_{\mathrm{MST}}$ (°) | +1.43° | 0.219 | +1.73° | 0.219 |

*Notes:* Δ denotes outlier minus inlier means within each component–depth cell before pairing; relative changes are expressed as percentages of the inlier mean. Exact ties (none here) would be dropped from $n$ for that comparison.

Alongside the deviation analysis, these findings suggest that variability in the calculated strain / stress tensors is not due to any single variable. Rather, the variability is due to the sampling of discretely different strain states, characterized by averages of select grain populations. This demonstrates that in samples with complex and fine-scale variations in microstructure and strain, the calculated solutions and uncertainties do not reflect the true volumetric average, even with infinitely accurate measurements. Based on this experimental setup, any one solution may be either tensile or compressive depending on the choice of measurements used (at least for the six-measurement method).

Building on the deviation analysis above, we next quantify how the choice of six orientations governs uncertainty amplification in the direct inversion of the design matrix $\boldsymbol{f}_j(\phi_i, \psi_i)$ (which we now write as $\boldsymbol{F}$ for brevity). Uncertainties in the measured strains are propagated to the sample strains through the information matrix,

$$\text{cov}(\boldsymbol{\varepsilon_j}) = \left(\boldsymbol{F}^{\top}\boldsymbol{W}\boldsymbol{F}\right)^{-1} \tag{21}$$

where $\boldsymbol{W} = \text{diag}\left(\frac{\mathbf{1}}{\boldsymbol{var}(e_i)}\right)$. The diagonal entries $[(\boldsymbol{F}^{\top}\boldsymbol{W}\boldsymbol{F})^{-1}]_{jj}$ are per-component variance multipliers. Larger values indicate that the chosen six orientations tend to amplify noise into component *j* more strongly (poorer geometric leverage on that component), while smaller values indicate better leverage.

To visualize how each orientation participates in conditioning of the information matrix $[(\boldsymbol{F}^{\top}\boldsymbol{W}\boldsymbol{F})^{-1}]_{jj}$, contour plots were constructed as follows. For each orientation $(\phi, \psi)$, we collect all solutions that contain that measurement and compute for each component $j$ the median of $[(\boldsymbol{F}^{\top}\boldsymbol{W}\boldsymbol{F})^{-1}]_{jj}$. Lower values mark orientations that tend to participate in well-constrained inversions. Higher values mark orientations that tend to participate in poorly constrained inversions. The results of this analysis for each location yielded similar results. Thus, only the results of the shallowest location are visualized in Figure 12.

This analysis shows how the observability of a particular strain / stress component along a particular measurement orientation is the dominant constraint on that component's solution. No particular variability in measurement uncertainty has any notable influence on noise amplification when inverting the information matrix. Rather, the choice of orientations strongly constrains the solution. The relative consistency of $[(\boldsymbol{F}^{\top}\boldsymbol{W}\boldsymbol{F})^{-1}]_{jj}$ for the shear components is a primary reason for the large variance of the solutions. It appears that optimal geometric conditioning of orientations is achieved by covering (i) the three sample axes: $S_{11}$, $S_{22}$, and $S_{33}$, and by (ii) choosing three interior angles with adequate spacing and a lack of redundant information due to symmetry. These findings reinforce intuitions regarding data collection.

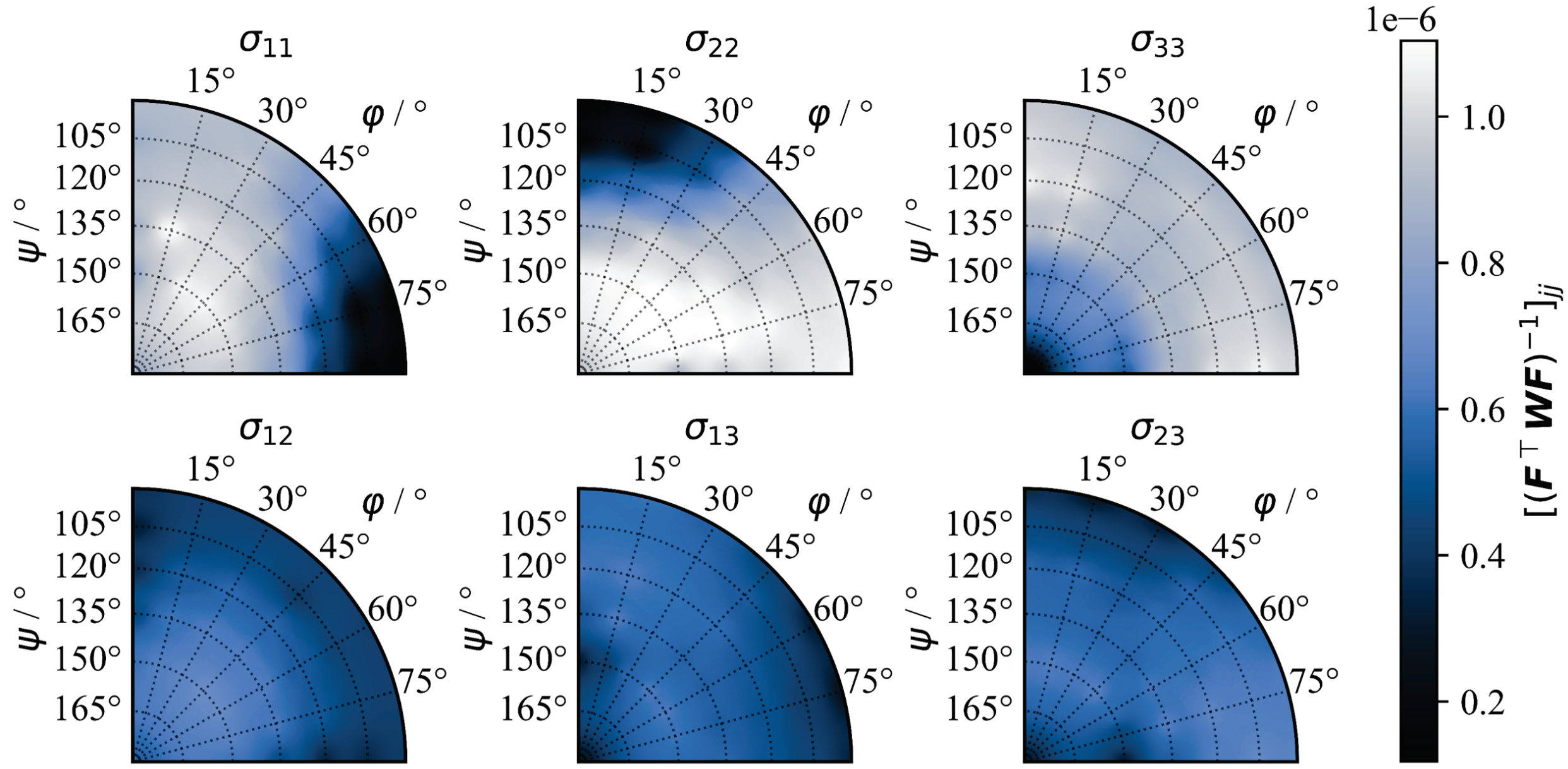

**Figure 12:** Per component conditioning metric $[(\boldsymbol{F}^{\top}\boldsymbol{W}\boldsymbol{F})^{-1}]_{jj}$, colored according to its median value across all solutions for a given orientation. These plots demonstrate that optimizing the observability of each component is a necessary conditioning metric, for which there are no preferred orientations for the shear components.

*Uncertainty analysis of least-squares method*

Incorporating thirty-six measurements into a least-squares procedure reduces the variance of solutions. One important caveat to acknowledge is that the propagated uncertainties of the least-squares estimator only describe the confidence of the fit. This is opposed to the confidence of the solution with respect to ground truth, which cannot be obtained by a singular least-squares estimation. As more measurements are included, the propagated uncertainty will converge to a small value relative to the individual measurement uncertainties. This phenomenon of uncertainty convergence is demonstrated in Figure 13 and Table S4. For example, the spread of nominal solutions and the average uncertainty for $\varepsilon_{11}$ fall below ± 100 με once about twenty-seven measurements are used. After thirty-six measurements are used, the uncertainty falls below ± 50 με.

As additional measurements are included in the least-squares fit, the propagated uncertainty on the fitted tensor decreases, as expected for an averaging estimator. However, the resulting uncertainty is only accurate if the per-measurement uncertainties fully describe the variability of the measurements about a single strain tensor. We test that assumption by examining the standardized residuals. These residuals are defined as

$$R_w = (e_{i,\text{obs}} - e_{i,\text{pred}})\sqrt{w_i} = \frac{e_{i,\text{obs}} - e_{i,\text{pred}}}{\mu(e_i)} \quad (22)$$

where $e_{i,\text{obs}}$ is the measured strain in the diffraction coordinate system, $e_{i,\text{pred}}$ is the corresponding strain predicted by inserting the least-squares sample tensor back into the strain transformation law at the measured $(\phi, \psi)$, and $\mu(e_i)$ is the one-standard-deviation uncertainty of the strains based on the peak-fit uncertainty.

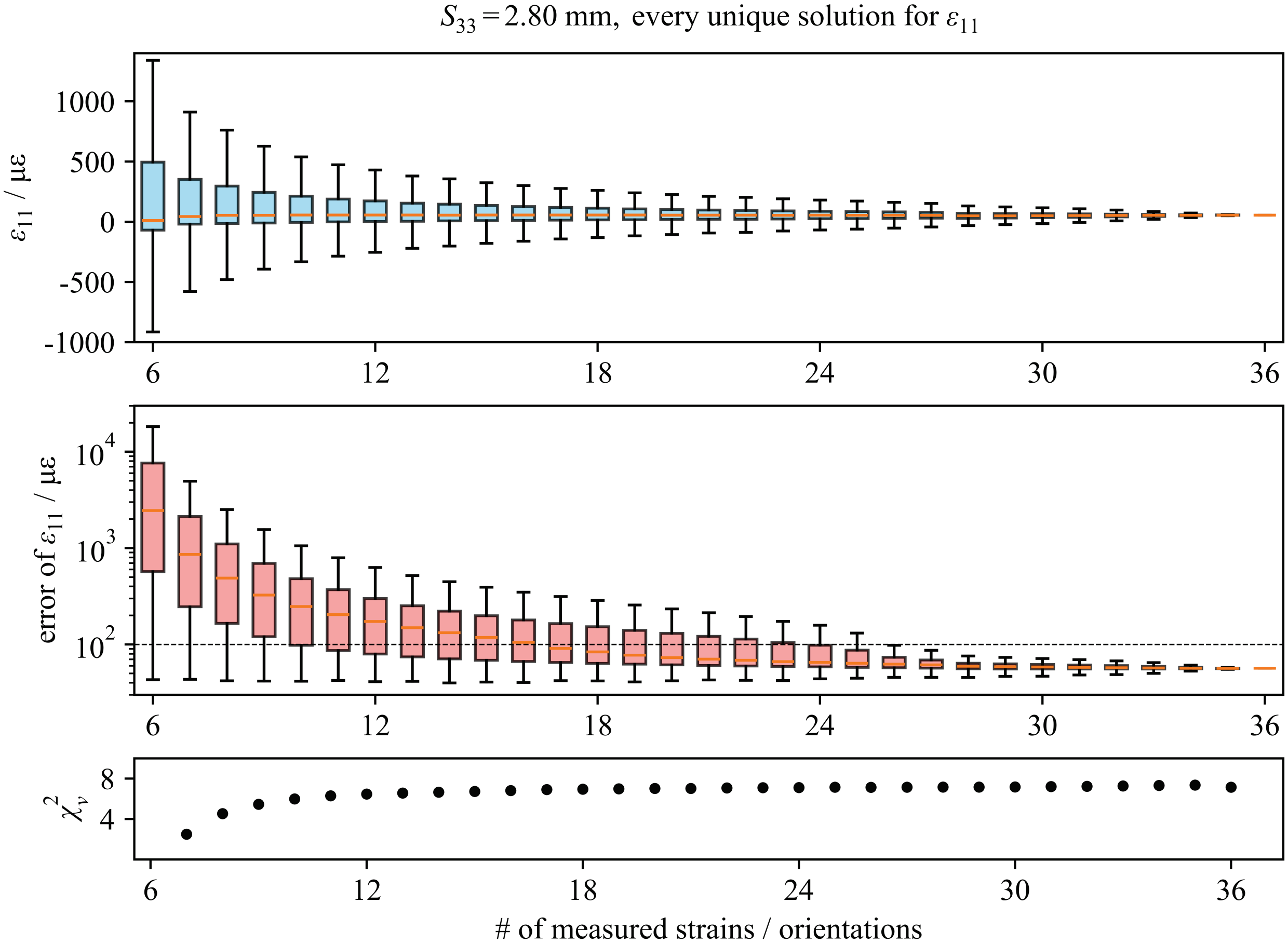

**Figure 13:** Boxplots of all calculable solutions and uncertainties for $\varepsilon_{11}$, computed from the least-squares method using orientations within the top volume. The median reduced chi-squared $\chi_v^2$ is shown on the bottom-most plot, demonstrating that as more measurements are included, the assumption of a single sampled strain-state breaks down. The boxplots are constructed from the medians (orange) and the 25th and 75th percentiles of data.

If the thirty-six measurements are consistent with a single strain tensor and the uncertainties are complete, then the residuals should have zero mean and unit variance. In that case, the goodness-of-fit statistic

$$\chi^2 = \sum_{i=1}^{N} R_{w,i}^2 \tag{23}$$

$$\chi_v^2 = \frac{\chi^2}{N-p} \tag{24}$$

should be consistent with a chi-squared distribution with $N - p = 30$ degrees of freedom.

Figure 14 shows the distributions of standardized residuals for each sampling location. Clearly, the spread of residuals is much broader than what would be predicted if the measurements were representative of the same strain state. Formalized chi-squared and t-tests were conducted at each location to assess the significance of high variance and systematic effects. The reduced chi-squared values range from 2.34 to 7.15, far above expected value of one. See table S4 for statistics at each depth. The t-tests do not reject a zero mean of residuals, indicating likely no significant systematic effects relative to the best-fit tensor. Overall, the residuals are centered near zero, but their spread is substantially broader than expected from the measurement uncertainties. Thus, while the least-squares tensor sits near the center of the data, the peak-fit uncertainties are too small to explain the observed residual scatter about a single-tensor. In standard statistical interpretation, such elevated values of reduced chi-squared would suggest that either the uncertainties are incomplete or that the underlying model does not describe the data.

We contend that the measurement uncertainties should be inflated until the reduced chi-squares equals 1. This is justified based on three assumptions, which are not always valid: (1) that the measurement errors (deviations from truth) behave randomly; (2) that the model is linear; and (3) that the model is correct and represents the underlying data (Andrae *et al.*, 2010). Often, these assumptions are not valid, as has been noted in the field of astronomy, where uncertainty estimation is paramount (Andrae, 2010; Andrae *et al.*, 2010). In addition, it is common for the model to be incorrect. However, our use case satisfies all three assumptions. First, the deviation analysis in Tables 2 and 3 shows no significant correlation between measurement variables and a solution's deviation from the average solution. Second, the strain-transformation equation is indeed a linear transformation of strains and trigonometric coefficients. When the first two assumptions are satisfied, the chi-squared statistics often reveal the inadequacy of simplified physical models via under or over-fitting of the data (i.e., $\chi_v^2 \gg 1$ or $\chi^2 < 1$, respectively). In our case, the strain transformation equations are a law, not a model, and they must be satisfied by any true macroscopic strain state. While each measurement does sample different populations of crystallites, giving rise to the observed deviations, it is justified to inflate uncertainties via the reduced chi-squared test because the results will be interpreted as macroscopic continuum strains (for structural integrity applications). Simply put, the uncertainties should be large enough to capture variations relative to the average strain state. In that context, measurement uncertainties may be inflated by a common factor until the standardized residuals pass the reduced chi-squared test. This means the inflated uncertainties propagated through the least-squares fit are more justified and conservative.

This interpretation is also supported by three prior observations. First, oscillatory behavior of the $\sin^2\psi$ plots indicates heterogeneity in the sampled strain-states. Second, the non-zero and solution-dependent deviation means $\mu_{D^{\varepsilon}_{ij}}$ in the six-measurement analysis suggest an orientation-dependent scatter. Third, the suppression of non-zero means from the strain deviations $\mu_{D^{\varepsilon}_{ij}}$ to zero means for the stress deviations $\mu_{D^{\sigma}_{ij}}$ implies a common, systematic effect (e.g. $d_0$).

Accordingly, a more complete uncertainty description recognizes that each solution to the strain tensor is offset from the true solution by a common bias $b_{ij}$, an orientation-dependent bias $\delta^{(k)}_{ij}$, and random noise:

$$\varepsilon^{(k)}_{ij} = \varepsilon^{true}_{ij} + b_{ij} + \delta^{(k)}_{ij} + noise^{(k)} \tag{25}$$

If these additional contributions ignored, the uncertainty propagated into the least-squares estimator no longer retains its interpretation as a one-standard deviation measure of variability (i.e. noise). Thus, the repeatability of the experiment is not ensured.

The common bias $b_{ij}$ is incalculable when the true strain is not known. However, the relative magnitude of the orientation-dependent biases may be estimated by introducing an additional scaling factor $C$ to the measurement uncertainties,

$$\mu_{e_{i,eff}} = C\mu_{e_i} \tag{26}$$

where $C$ is chosen such that the reduced chi-squared satisfies $\chi^2_\nu = 1$ (see Table S4). Depending on location, the required scale factor ranges from $C$ = 1.53 to 2.67. The spread indicates a location (microstructure) dependent effect. More specifically, the intensity of small-scale microstructural gradients is greatest for the top-most volume $S_{33}$ = 2.8 mm. This top-most volume has undergone less thermal cycling and recrystallization compared to the locations below, a difference noted previously for the results of Table 3. This scaling does not alter the nominal least-squares solution. The distributions of standardized residuals, before and after uncertainty inflation, are shown alongside each other in Figure 14. The results of the least-squares estimation after uncertainty inflation are given in Figure 15. The uncertainty intervals are similar to the three measurement results, which suggests that the uncertainties for the latter – which is the common method – are underestimated.

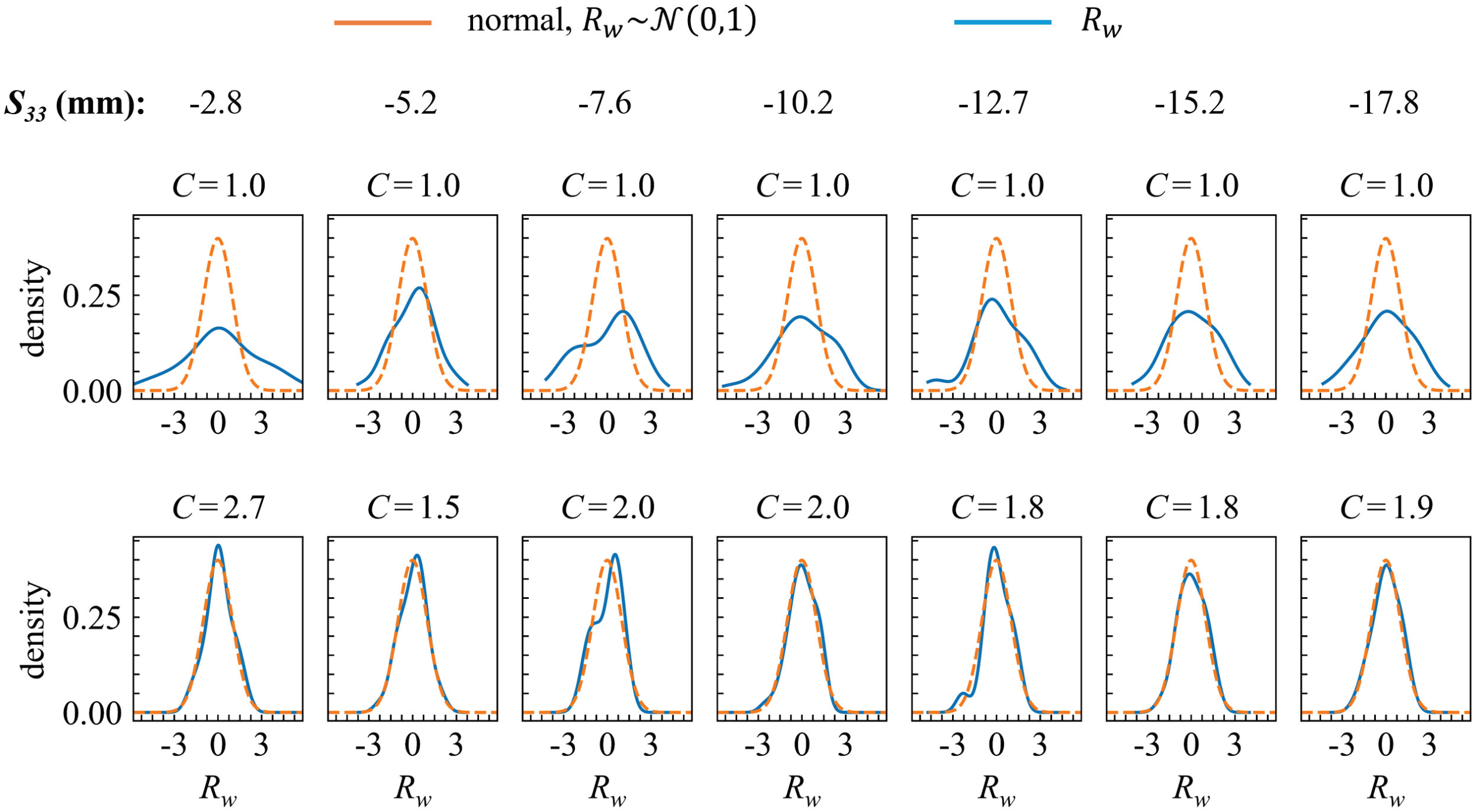

**Figure 14:** Distributions of residuals expressed as continuous functions (KDEs), If the measurements represent only random noise about a singular strain state, the weighted residuals should model a normal distribution, $R_w \sim \mathcal{N}(0,1)$. The top row shows that the measured uncertainties yield weighted residuals inconsistent with random noise, whereas the bottom row shows that inflating the uncertainties produces residuals consistent with random noise. This suggests that inflating the measured uncertainties is required to represent the experimental scatter.

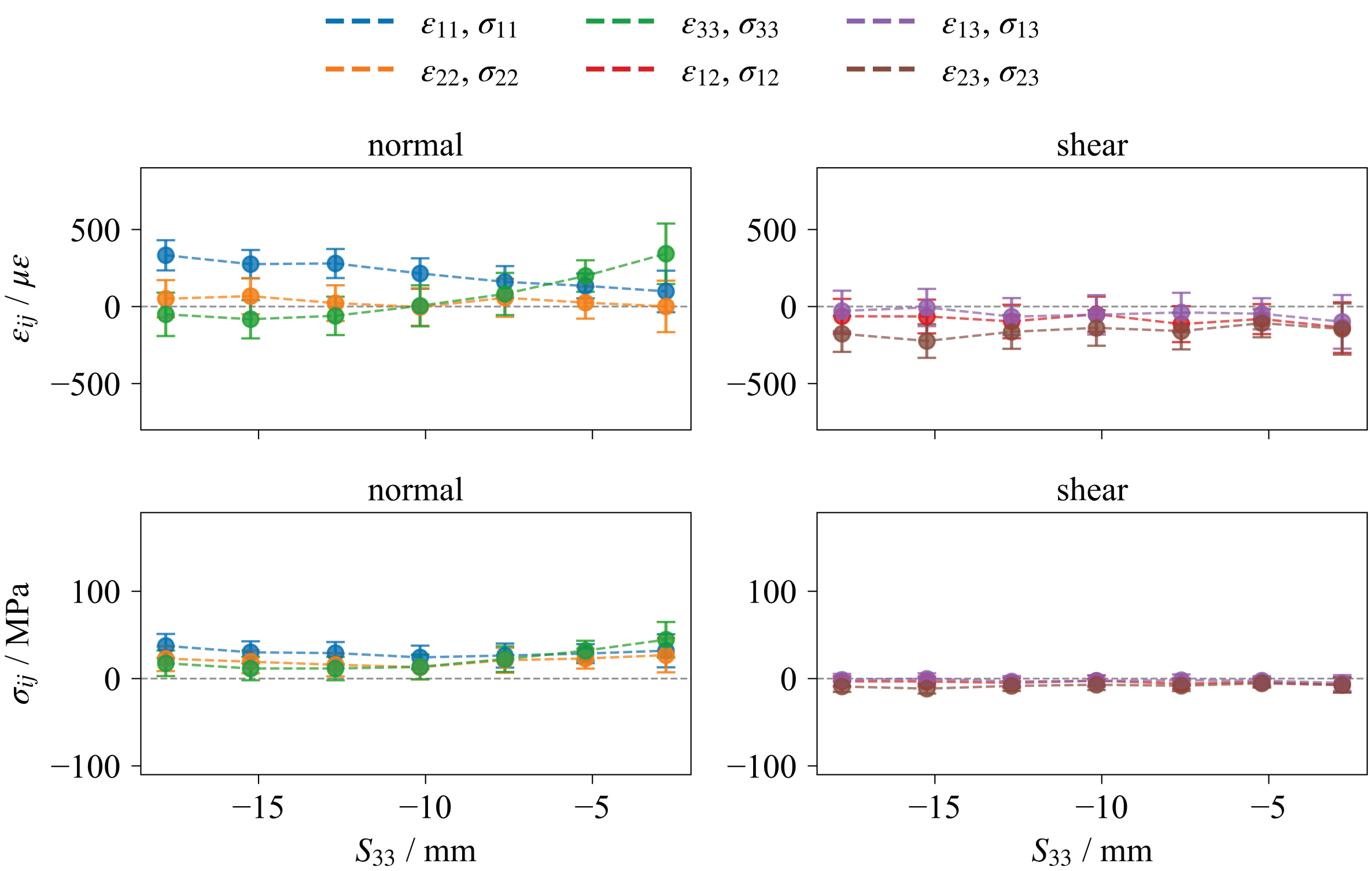

**Figure 15:** Results of the strain / stress solutions and their errors after inflating measurement uncertainties to account for hidden biases. The uncertainties are two to three times larger than those of the un-inflated least-squares estimator, yet they are similar to the three-measurement solutions. This implies that the three-measurement uncertainties are too small since the measurement biases could not be evaluated. In addition, it is clear that the uncertainty scales with the amount of orientation dependent scatter, as seen for the top-most location ($z$ = -2.8 mm).

*Suggestions for ideal count times.*

A key implication of this work is that an ideal diffraction measurement is not defined solely by precision, but by whether its reported uncertainty meaningfully represents the variability of the underlying strain state. In conventional practice, a minimal set of orientations is measured and count times are increased to reduce peak-fit uncertainty. As shown here, this approach can produce uncertainties that are artificially small and unrepresentative of the true experimental scatter when measurements sample different strain states.

Instead, a more appropriate strategy is to measure redundant orientations and evaluate consistency through the strain transformation law. In this framework, least-squares estimation reduces the potential systematic effects of orientation-dependent variability, while statistical tests such as the reduced chi-squared provide a means to assess whether the propagated uncertainties remain valid. When they do not, the uncertainties can be adjusted to reflect the effective variability of the measurements. This approach reframes uncertainty as a quantity that must be validated against physical consistency, rather than minimized through counting statistics alone.

However, such strategies introduce practical constraints, as measurement time is limited and cannot be arbitrarily increased across many orientations. This motivates the need for optimized measurement strategies that balance orientation coverage, count time, and the reliability of the resulting uncertainty estimates. Thus, the following strategy is proposed.

For the sake of argument, it is assumed that peak-fit uncertainty is proportional to count time based on counting statistics of random noise,

$$\mu \propto \frac{1}{\sqrt{N}} \tag{27}$$

where $N$ is the total number of counts, which increases linearly with time. Therefore, uncertainty should decrease with time according to the following relationship:

$$\mu = \mu_0 \sqrt{\frac{t_0}{t}} \tag{28}$$

where $t_0$ is the original count time with associated uncertainty $\mu_0$, and $t$ is the new count time. If uncertainties $\mu$ should decrease by a certain factor $f$ from their original value $\mu_0$, then the new time is given by:

$$t_{new} = \frac{t_0}{f^2} \tag{29}$$

based substituting the relation $\mu(t_{new}) = f\mu_0$ into Eq. 28. Since the uncertainties needed to be inflated by at least a factor of 1.5, the time for each measurement could have been reduced from 10 minutes to 4.4 minutes (assuming the same measurements are taken).

It is possible to define an optimization scheme where one measures a large number of strains (e.g. 36) at various angles $(\phi_i, \psi_i)$ within a representative microstructure of the component. Based on these measurements, an objective is set based on the desired count time or uncertainty in the best-fit strain-tensor. For example, the desired maximum uncertainty in the best-fit strain tensor can be set to some reasonable fraction of the material's yield strength (e.g. for annealed AA6061, approximately 100 με). A global minimization strategy is performed for a given objective function, such as, "Given a target maximum uncertainty in the best-fit strain tensor (e.g., max $\mu_{\varepsilon_{ij}} \leq 100$ με), what is the minimum number of orientations and the minimum count time per orientation needed?" Based on this objective, an algorithm will set a stop limit during an iterative process whereby the algorithm will loop through the appropriate number of measured orientations $(\phi_i, \psi_i)$ and associated count times to achieve this metric. Such a process would make full-tensor calculations more confident and economical.

**Conclusions**

This work demonstrates that the interpretation of propagated uncertainty in diffraction-based residual strain analysis depends critically on both the calculation pathway and the underlying assumption of a single, representative strain state. In the additively manufactured component studied here, fine-scale gradients in texture, plastic strain, and elastic strain lead to orientation-dependent sampling of distinct strain states. As a result, one-standard-deviation peak-fit uncertainties do not consistently represent random noise about a common mean once propagated to the solution.

For the six-measurement direct inversion, propagated uncertainties remain meaningful and generally conservative with respect to the empirical spread of solutions. In contrast, the thirty-six-measurement least-squares approach reduces variance through averaging, but produces uncertainty intervals that underestimate the true variability of the measurements. In this case, the reported uncertainties reflect overconfidence in the fit rather than capturing the physical scatter of the data.

These results show that uncertainty underestimation in diffraction is not solely due to instrumental or fitting limitations, but is strongly influenced by intrinsic, sample-dependent effects. Consequently, agreement with the strain transformation law provides a powerful physical check on the validity of propagated uncertainties. An important implication is that oversampling does not inherently improve the reliability of uncertainty estimates. While least-squares methods stabilize the solution, they may obscure physically meaningful variability when the assumption of a singular strain state is compromised. For structural integrity applications, this distinction is critical: a low-variance estimate is not necessarily a conservative one.

Finally, this work provides a practical pathway for improving uncertainty estimates through oversampling the number of measured orientations and statistical consistency checks. By evaluating residuals and enforcing agreement under the strain transformation law, measurement uncertainties can be adjusted to better reflect the effective variability of the experiment. Reliable residual stress assessment in heterogeneous materials therefore requires not only precise measurements, but also explicit validation of the assumptions underlying uncertainty propagation.

We suggest that more accurate strain estimates with more realistic uncertainties may be achieved by oversampling orientations with reduced count time per measurement, rather than by acquiring a minimal set of orientations with longer count times.

Future work should repeat these analyses on an ideal reference standard, such as a ring-and-core, across multiple beamlines in a round-robin style experiment. This will improve calibration and reporting of fundamental uncertainties for neutron diffraction beamlines meant for strain determination.

**Acknowledgments**

This work makes use of perceptually uniform colormaps developed by Fabio Crameri (Crameri *et al.*, 2020). This research used resources at the High Flux Isotope Reactor, a DOE Office of Science User Facility operated by the Oak Ridge National Laboratory. The beam time was allocated to the HIDRA instrument on proposal number IPTS-33702.1. The authors are thankful for Rob Patterson's help creating the AFSD samples. The authors also thank Dr. Benjamin Wing for assistance at the High Flux Isotope Reactor.

**Funding**

This material is based upon work supported by DEVCOM Army Research Laboratory Cooperative Agreement W911NF21200.

**CrediT Statement**

Cole Franz: Conceptualization, Formal Analysis, Investigation, Writing – original draft; Michael B. Prime: Conceptualization, Investigation, Writing – review and editing. Jeffrey Bunn: Investigation, Writing – review and editing. Andrew Payzant: Investigation, Writing – review and editing. Katharine Page: Funding acquisition, Investigation, Writing – review and editing.

**Statements and Declarations**

Not Applicable

**Supplemental Information**

*AFSD Deposition Process*

Extruded AA6061-T6 bar stock (12.7 x 12.7 mm$^2$) was deposited onto a rolled AA6061-T6 baseplate (152.4 x 152.4 x 9.5 mm$^3$) using an L3 discrete feed AFSD machine (MELD Mfg., Christiansburg, VA, USA). In total, 11 unidirectional layers were deposited. The waiting period between each deposition end and start was 1 minute, with 3 minute periods occurring every three layers for feedstock reloading. The tool was constructed primarily of Be-Cu and had an H13 face. On the tool face are four 0.635 mm nub features, and underneath the face is a K-type thermocouple used to control spindle speed through a temperature-controlled feedback loop. During deposition, the spindle speed is constantly adjusted such that the tool thermocouple maintains a setpoint temperature of 425 °C. The process parameters are given in Table S1.

**Table S1**: AFSD process parameters.

| Parameter | Value | Units |
|---|---|---|
| Layer Height ($h$) | 2.54 | $mm$ |
| Rotational Speed ($\omega$) | 160 | $rpm$ |
| Velocity ($V$) | 3.8 | $\frac{mm}{sec}$ |
| Feed Rate (FR) | 2.9 | $\frac{mm}{sec}$ |
| Temp. Setpoint | 420 | °C |
| Tool Nub Height | 2.3 | mm |
| Bar Stock | 12.7 x 12.7 | $mm^2$ |

*Neutron Diffraction*

It is assumed that a macro-scale strain-free reference was achieved by mechanical relief. For this purpose, a 5 mm diameter pillar was sectioned from an identical build, covering the same volumes as those sampled in the un-sectioned component. Photographs of the $d_0$ texture measurement setup and of the sectioned component are shown in Figure S1.

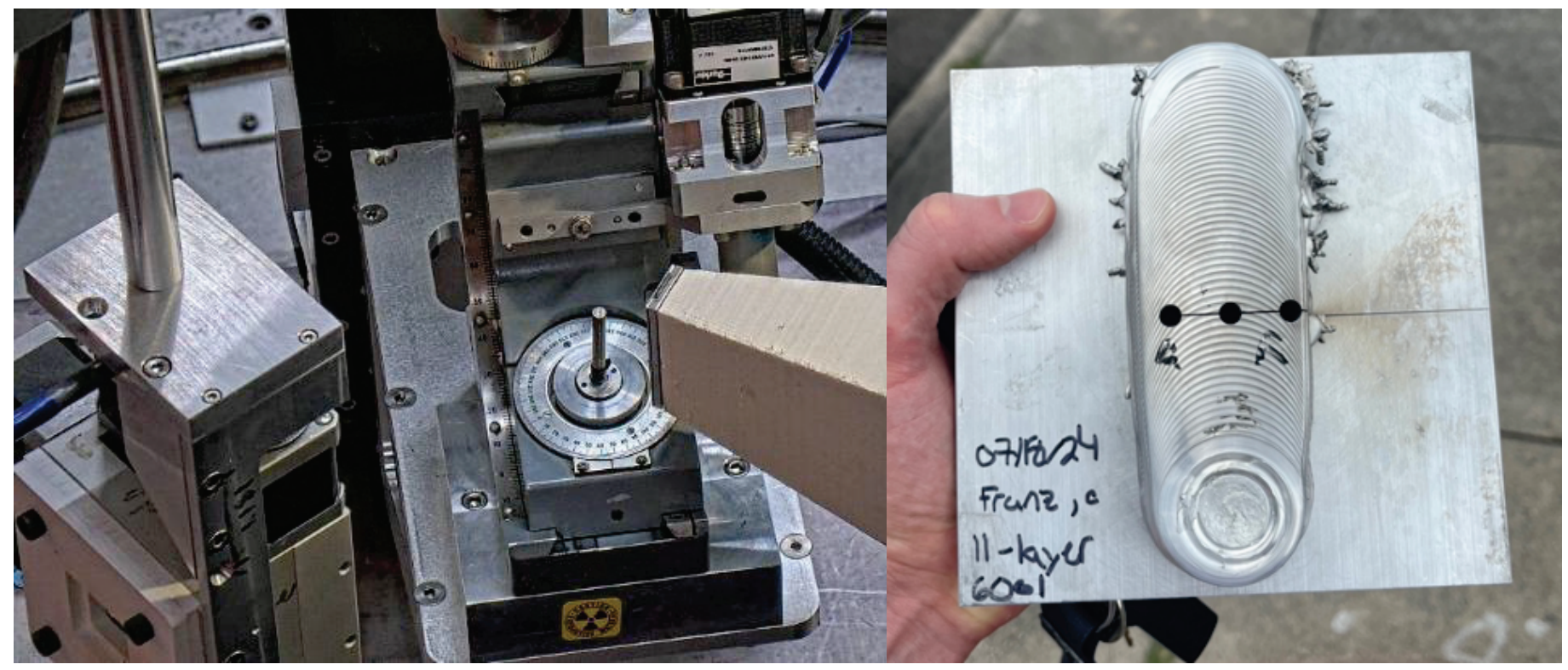

**Figure S1:** Photographic images of the texture data collection within the $d_0$ cylinder (left) and of the sectioned component after pillar extraction (right) (only the central pillar was used in this work, following the locations measured in the full component).

The bulk stiffness tensor was computed based on single crystal Al properties and the reconstructed orientation distribution functions (ODFs).

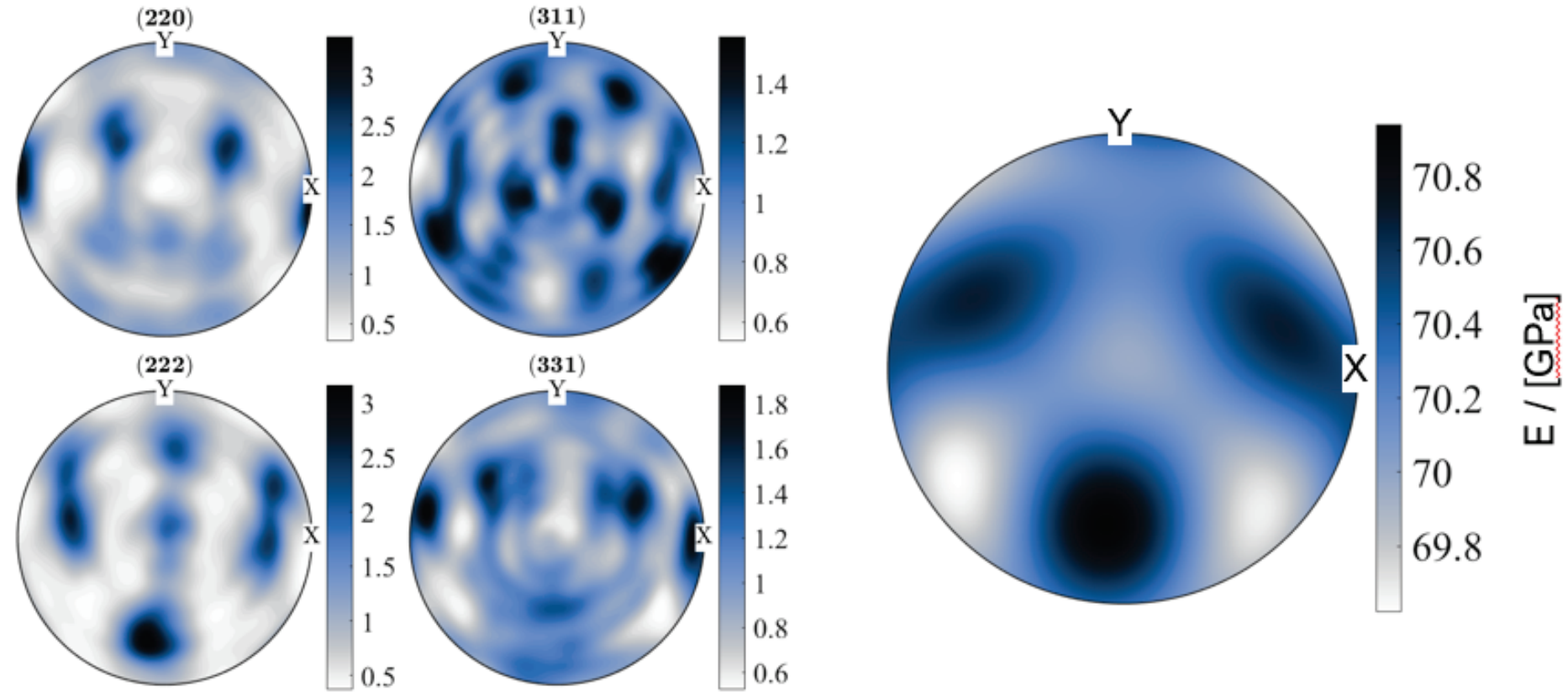

**Figure S2:** Reconstructed pole figures and bulk-scale elastic modulus based on texture data collected in the top-most volume. No more than 1 GPa variation in stiffness is predicted on the bulk-scale, which warrants use of isotropic elasticity to calculate stresses within the measurement volumes.

*Lattice parameter variation*

To estimate the theoretical bounds of the unstrained lattice parameter in AA6061, we employed a thermodynamically grounded model based on modifications of Vegard's law with Redlich-Kister Polynomials, which linearly interpolates the lattice parameter from the atomic fractions of constituent elements assuming ideal substitutional mixing:

$$a_0 = \sum_i x_i a_i \tag{S1}$$

However, in systems like Al-Si-Mg or Al-Mg-Cu where binary interactions are significant, the Redlich-Kister polynomial formulation offers a more accurate description. This includes pairwise interaction terms between solutes *i* and *j*, weighted by their mole fractions:

$$a_0 = \sum_i x_i a_i + \sum_{i=1}^{n} \sum_{j=i+1}^{n} x_i x_j \sum_{n=0}^{k} L_n^{(ij)}(T)\left(x_i - x_j\right)^n \tag{S2}$$

Here, $L_n^{(ij)}(T)$ are interaction coefficients that depend on temperature. The temperature dependence was included based on coefficients reported by Shin et al. (2017) for Al-Mg, Al-Si, and Al-Cu binary systems (Shin *et al.*, 2017). We implemented these expressions, using temperature-dependent lattice parameters for pure Al, Mg, Si, and Cu and evaluated the resulting lattice parameter of the alloy at room temperature.

To convert this lattice parameter to the expected interplanar spacing for the (311) reflection, we used the cubic lattice relation:

$$d_{311} = \frac{a}{\sqrt{h^2+k^2+l^2}} = \frac{a}{\sqrt{11}} \tag{S3}$$

Two ternary systems were considered: Al-Si-Mg and Al-Mg-Cu. For each, the full range of solute concentrations permitted by solubility limits or compositional bounds of AA6061 was explored. This included atomic fractions of Mg up to 1.35 at.%, Si up to 0.78 at.%, and Cu up to 0.17 at.%—values representative of the upper limits for supersaturation.

To further refine the model, we considered that some fraction of Mg and Si may be sequestered into $Mg_2Si$ precipitates during solidification or aging. We treated the total alloy composition as a mixture of matrix and precipitate contributions and solved for matrix solute contents $x_{i,m}$,

$$x_{i,\mathrm{total}} = V_p x_{i,p} + \left(1 - V_p\right) x_{i,m} \tag{S4}$$

where $V_p$ is the precipitate volume fraction and $x_{i,p}$ is the atomic fraction of element $i$ in the precipitate. Assuming a stoichiometry of $Mg_2Si$, we predicted the maximum volume fraction of precipitate that could form before fully depleting one solute. For AA6061 chemistry, Mg was exhausted first, yielding a maximum precipitate volume fraction of 2.01%. Beyond this, any excess Si was assumed to form separate phases or remain undetected. Based on complete supersaturation and precipitation, the maximum deviation in (311) interplanar spacing is presented in Figure S3.

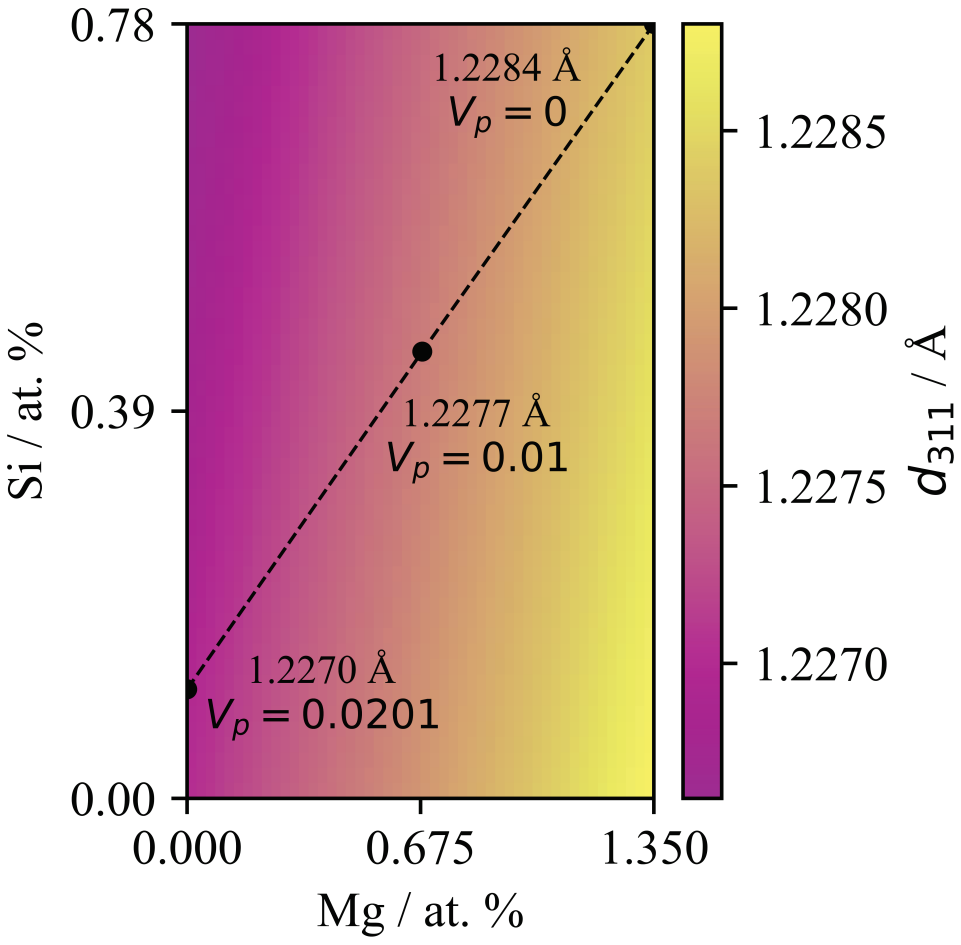

**Figure S3:** Predicted variation in the (311) interplanar spacing for an Al-Si-Mg alloy system at room temperature. The heatmap represents a range of matrix-level Si and Mg concentrations within the composition limits of AA6061. The dashed line connects three physically relevant points corresponding to different assumed precipitate volume fractions $V_p$, with their associated $d_{311}$ values annotated. These points represent the maximum solute content in solid solution (no precipitation), partial precipitation (1%), and complete precipitation of Mg (2.01%)—after which no further reduction in matrix Mg is possible.

*Six-measurement deviations correlations*

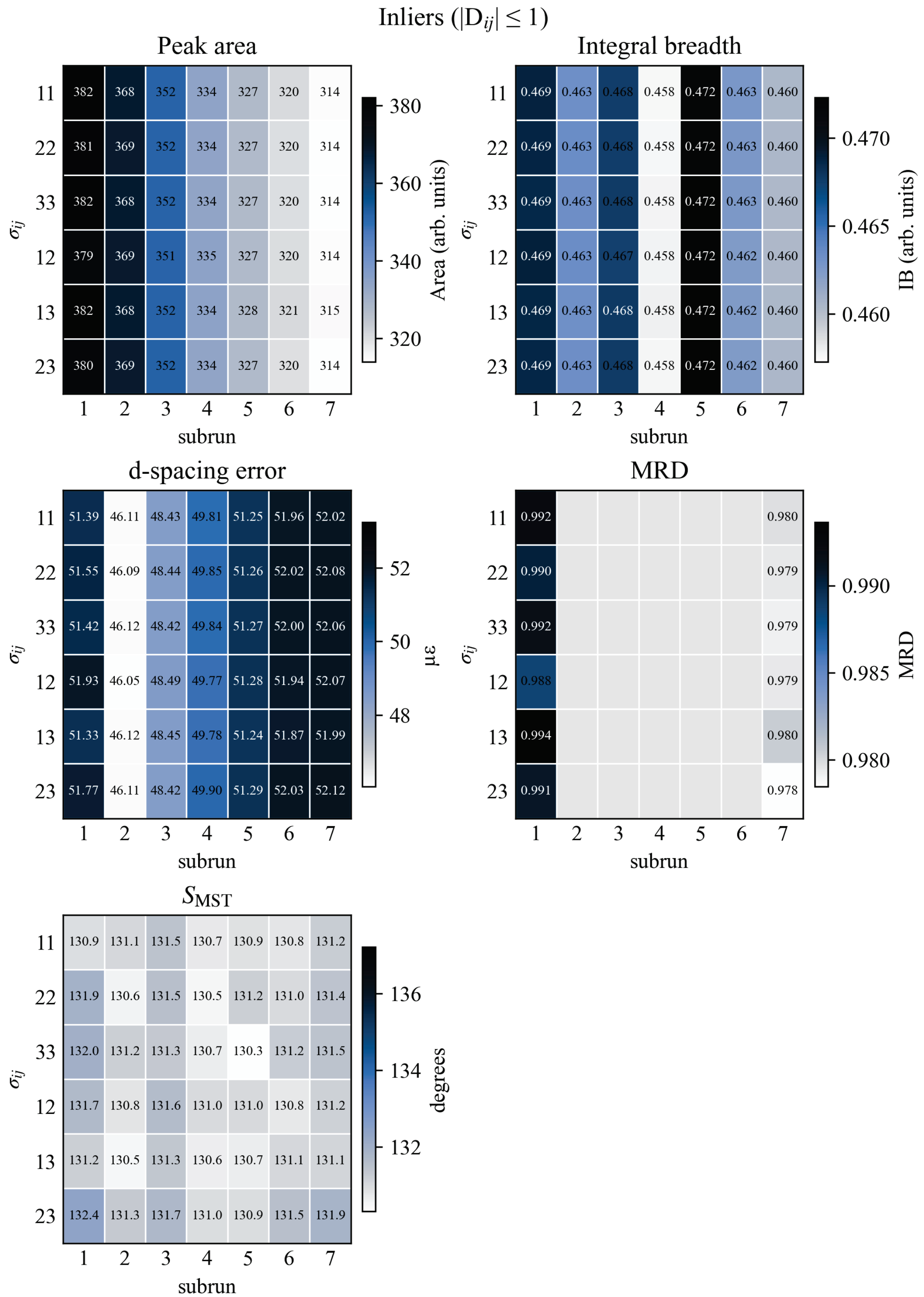

**Figure S4:** Average metrics of deviation inliers for the six-component solutions to stress.

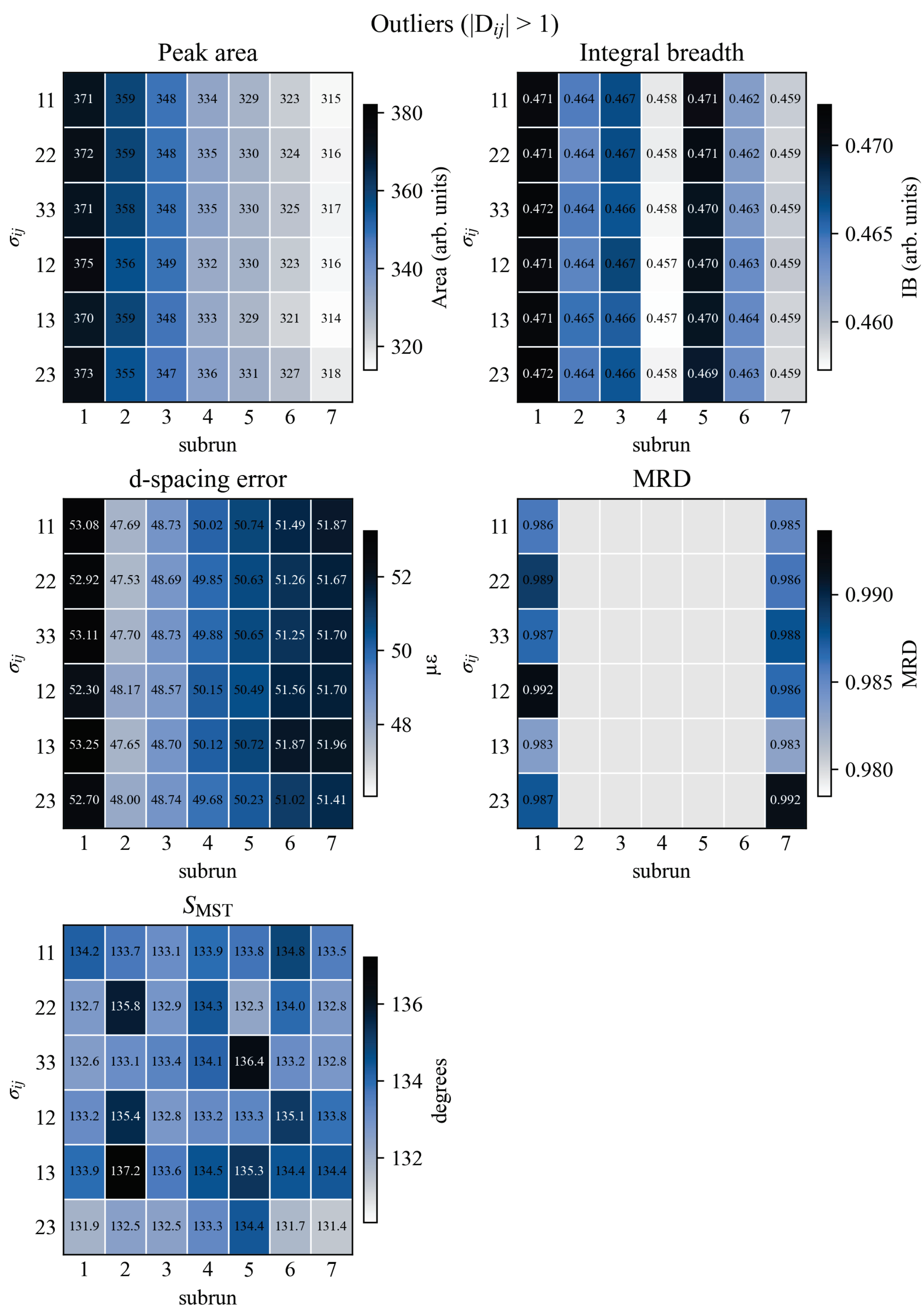

**Figure S5:** Average metrics of deviation outliers for the six-measurement solutions to stress.

*Convergence of least-squares uncertainties*

**Table S2:** Number of orientations of strain measurements within a set for the standard deviation of the nominal solutions and the mean of the error to be less than or equal to ± 100 με.

| $S_{33}$ (mm) | $\varepsilon_{11}$ | $\varepsilon_{22}$ | $\varepsilon_{33}$ | $\varepsilon_{12}$ | $\varepsilon_{23}$ | $\varepsilon_{13}$ |
|---|---|---|---|---|---|---|
| -2.8 | 20 | 29 | 28 | 30 | 30 | 25 |
| -5.21 | 19 | 29 | 24 | 25 | 27 | 22 |
| -7.62 | 20 | 33 | 26 | 24 | 29 | 25 |
| -10.16 | 19 | 32 | 26 | 25 | 29 | 25 |
| -12.7 | 20 | 32 | 27 | 26 | 29 | 26 |
| -15.2 | 21 | 34 | 27 | 27 | 31 | 27 |
| -17.8 | 20 | 33 | 28 | 25 | 34 | 26 |

*Significance of least-squares residuals*

To formally assess whether the residuals satisfy the assumptions of zero mean and unit variance, we performed t-tests for bias and $\chi^2$-tests for variance at the 95% confidence level using the weighted residuals computed prior to adding any extra uncertainty. The results for each location are presented in Table S3.

Across all measurement locations, the mean residuals $\mu_{R_w}$ were small, and the t-tests showed that none exceeded the critical threshold $t_{0.975,30} = 2.042$. Statistically, this means we cannot reject the null hypothesis that the residual means are zero at the 95% confidence level. However, this does not imply that the strain transformation law is fully valid or that the measurements are free of additional systematic effects. Rather, it indicates that the best-fit tensor lies near the center of the measured data. In contrast, the $\chi^2$-tests for the residual variances rejected the null hypothesis of unit variance at all locations, with $\chi^2$ statistics far exceeding the acceptance interval [17, 47]. This demonstrates that the reported measurement uncertainties alone are insufficient: the residuals exhibit much broader scatter than would be expected from the nominal peak-fit uncertainties. Physically, this implies that the propagated measurement uncertainties underestimate the true experimental scatter, motivating the introduction of an additional uncertainty scale factor.

**Table S3:** Results of bias and variance significance tests using t-tests, $\chi^2$ tests, and 95% confidence intervals for the least-squares estimator without uncertainty inflation (i.e., using only the reported peak-fit uncertainties as the measurement uncertainties).

| $S_{33}$ (mm) | $\mu_{R_w}$ | $\sigma_{R_w}$ | $t$ | $t^*_{0.975,30}$ | bias test | $\chi^2_v$ | $\left[\chi v^2_{0.005,30}, \chi v^2_{0.975,30}\right.$ | var. test |
|---|---|---|---|---|---|---|---|---|
| -2.8 | 0.24 | 2.46 | 0.59 | | pass | 7.1 | | fail |
| -5.2 | 0.07 | 1.42 | 0.28 | | pass | 2.3 | | fail |
| -7.6 | 0.05 | 1.85 | 0.15 | | pass | 4.0 | | fail |
| -10.2 | 0.05 | 1.85 | 0.17 | 2.04 | pass | 4.0 | [0.6, 1.6] | fail |
| -12.7 | 0.05 | 1.67 | 0.19 | | pass | 3.3 | | fail |
| -15.2 | 0.09 | 1.62 | 0.32 | | pass | 3.1 | | fail |
| -17.8 | 0.07 | 1.72 | 0.26 | | pass | 3.5 | | fail |

*Notes:* All t-tests assume a 95% confidence interval and degrees of freedom $v = N - p = 36 - 6 = 30$.

**Table S4:** Results of bias and variance significance tests using t-tests, $\chi^2$ tests, and 95% confidence intervals for the least-squares estimator with uncertainty amplification (i.e., each measurement uncertainty is inflated by constant factor *C*).

| $S_{33}$ (mm) | $\mu_{R_w}$ | $\sigma_{R_w}$ | $t$ | $t^*_{0.975,30}$ | bias test | $\chi^2_v$ | $\left[\chi^2_{0.005,30}, \chi^2_{0.975,30}\right]$ | var. test |
|---|---|---|---|---|---|---|---|---|
| -2.8 | 0.09 | 0.9 | 0.59 | | pass | 1 | | pass |
| -5.21 | 0.04 | 0.9 | 0.28 | | pass | 1 | | pass |
| -7.62 | 0.02 | 0.9 | 0.15 | | pass | 1 | | pass |
| -10.16 | 0.03 | 0.9 | 0.17 | 2.04 | pass | 1 | [0.6, 1.6] | pass |
| -12.7 | 0.03 | 0.9 | 0.19 | | pass | 1 | | pass |
| -15.24 | 0.05 | 0.9 | 0.32 | | pass | 1 | | pass |
| -17.78 | 0.04 | 0.9 | 0.26 | | pass | 1 | | pass |

*Notes:* All t-tests assume a 95% confidence interval and degrees of freedom $v = N - p = 36 - 6 = 30$.

A common multiplicative scale factor $C$ was therefore applied to the quoted one-standard-deviation measurement uncertainties until the reduced chi-squared of the weighted residuals was equal to unity. The resulting post-calibration residual statistics are reported in Table S4. These values confirm the expected outcome of the calibration procedure and provide the revised uncertainty model used to propagate the least-squares tensor uncertainties.

*The effect of counting statistics on uncertainty.*

To evaluate whether excess residual scatter is correlated to counting statistics, we computed Spearman's rank correlation coefficient between the absolute least-squares residuals $|R_w|$ and the measured diffraction peak heights for each location. Spearman's $\rho$ measures the strength of a monotonic relationship between two variables as

$$\rho = 1 - \frac{6\sum d_i^2}{n(n^2-1)} \quad \text{(S5)}$$

where $d_i$ is the difference between the ranks of the two variables for the $i$-th observation, and $n$ is the number of paired observations. Here, $\rho = +1$ indicates a perfect positive monotonic correlation, $\rho = -1$ a perfect negative monotonic correlation, and $\rho = 0$ no monotonic relationship.

For each location, the null hypothesis $H_0\colon \rho = 0$ assumes no correlation between peak height and residual magnitude, while the alternative hypothesis $H_1\colon \rho \neq 0$ assumes a correlation exists. To assess statistical significance, $\rho$ was converted to a $t$-statistic using

$$t = \rho\sqrt{\frac{n-2}{1-\rho^2}} \quad \text{(S6)}$$

which follows a $t$-distribution with $n-2$ degrees of freedom under $H_0$. The corresponding two-sided p-value was computed as

$$p = 2[1 - F_t(|t|; n-2)] \quad \text{(S7)}$$

where $F_t$ is the cumulative distribution function of the $t$-distribution. Small p-values indicate that a correlation at least as extreme as the observed $\rho$ would be unlikely under $H_0$, and $p < 0.05$ was taken as evidence to reject $H_0$ at the 95% confidence level.

Results showed that only location 2 exhibited a statistically significant correlation ($\rho = -0.356$, $p = 0.0328$), indicating a weak negative monotonic relationship between peak height and residual magnitude at this location. All other subruns exhibited $p > 0.05$, implying no statistically significant correlation across most locations. Physically, this suggests that, except for one isolated region, residual magnitude is not systematically influenced by diffraction peak intensity. Thus, counting statistics was not the dominant experimental driver for systematic effects on the strain / stress solutions and uncertainties.

**References**

Akrivos, V., Wimpory, R. C., Hofmann, M., Stewart, B., Muránsky, O., Smith, M. C. & Bouchard, J. (2020). *J. Appl. Crystallogr.* **53**, 1181–1194.

Anderson, K., Weritz, J. & Kaufman, J. G. (2019). 6061 and Alclad 6061 Materials Park: ASM International.

Andrae, R. (2010).

Andrae, R., Schulze-Hartung, T. & Melchior, P. (2010).

Bachmann, F., Hielscher, R. & Schaeben, H. (2010). *Solid State Phenomena* **160**, 63–68.

Beghini, M. & Grossi, T. (2024). *Exp. Mech.* **64**, 851–874.

Beghini, M., Grossi, T., Prime, M. B. & Santus, C. (2023). *Exp. Mech.* **63**, 495–516.

Bunn, J. R., Fancher, C. M., Payzant, E. A., Cornwell, P. A., Bailey, W. B. & Gregory, R. (2023). *Review of Scientific Instruments* **94**, https://doi.org/10.1063/5.0122250.

Bunn, J. R., Penumadu, D., Lou, X. & Hubbard, C. R. (2014). *Metall. Mater. Trans. A Phys. Metall. Mater. Sci.* **45**, 3806–3813.

Crameri, F., Shephard, G. E. & Heron, P. J. (2020). *Nat. Commun.* **11**, 5444.

Fancher, C. M., Bunn, J. R., Bilheux, J., Zhou, W., Whitfield, R. E., Borreguero, J. & Peterson, P. F. (2021). *J. Appl. Crystallogr.* **54**, 1886–1893.

Hielscher, R. & Schaeben, H. (2008). *J. Appl. Crystallogr.* **41**, 1024–1037.

Holden, T. M., Traore, Y., James, J., Kelleher, J. & Bouchard, P. J. (2015). *J. Appl. Crystallogr.* **48**, 582–584.

Hutchings, M. T., Withers, P., Holden, T. M. & Lorentzen, T. (2005). *Taylor & Francis* 149–202.

JCGM GUM-6:2020 Guide to the expression of uncertainty in measurement-Part 6: Developing and using measurement models Guide pour l'expression de l'incertitude de mesure-Partie 6: Élaboration et utilisation des modèles de mesure (2020).

Metz, P. C., Arwood, Z., Franz, C., Heikkenen, E., Chawla, V., Babu, S. S., Penumadu, D. & Page, K. (2023). *Materialia (Oxf).* **30**, 101799.

Metz, P. C., Franz, C., Kincaid, J., Schmitz, T., Lass, E. A., Babu, S. S. & Page, K. (2024). *Addit. Manuf.* **81**, 103989.

Metz, P. C., Miller, L., Kincaid, J., Charles, E., Wood, A. T., Sims, Z. C., Drewry, S., Houston, A., Bunn, J. R., Compton, B., Schmitz, T., Lass, E. A., Penumadu, D. & Page, K. Through-Thickness Microstructure and Residual Stress Distributions in Additive Friction Stir Deposited Aluminum 7075.

Noyan, I. C. & Cohen, J. B. (1987). Residual stress : measurement by diffraction and interpretation New York: Springer-Verlag New York.

Prime, M. B., DeWald, A. T., Hill, M. R., Clausen, B. & Tran, M. (2014). *Eng. Fract. Mech.* **116**, 158–171.

Prime, M. B. & Hill, M. R. (2005). *J. Eng. Mater. Technol.* **128**, 175–185.

Rutherford, B. A., Avery, D. Z., Phillips, B. J., Rao, H. M., Doherty, K. J., Allison, P. G., Brewer, L. N. & Brian Jordon, J. (2020). *Metals (Basel).* **10**, 1–17.

Şeren, M. H., Pagan, D. C., Noyan, I. C. & Keckes, J. (2023). *J. Appl. Crystallogr.* **56**, 1144–1167.

Shin, D., Roy, S., Watkins, T. R. & Shyam, A. (2017). *Comput. Mater. Sci.* **138**, 149–159.

Winholtz, R. A. & Cohen, J. B. Generalised Least-squares Determination of Triaxial Stress States by X-ray Diffraction and the Associated Errors.

Withers, P. J. & Bhadeshia, H. K. D. H. (2001*a*). *Materials Science and Technology* **17**, 355–365.

Withers, P. J. & Bhadeshia, H. K. D. H. (2001*b*). *Materials Science and Technology* **17**, 366–375.

Zhu, N., Avery, D. Z., Chen, Y., An, K., Jordon, J. B., Allison, P. G. & Brewer, L. N. (2023). *J. Mater. Eng. Perform.* **32**, 5535–5544.